%% file: ms.tex
\documentclass[journal, 12pt]{IEEEtran}
\pdfoutput=1
\usepackage{lineno}
\usepackage[noadjust]{cite}
\usepackage{algpseudocode} 

\usepackage[]{graphicx}
\DeclareGraphicsExtensions{.pdf}

\graphicspath{{figures/pdf/}}

\usepackage{setspace}
\usepackage[cmex10]{amsmath}
\usepackage{amssymb}

\ifCLASSOPTIONcompsoc
 \usepackage[caption=false,font=normalsize,labelfont=sf,textfont=sf]{subfig}
\else
 \usepackage[caption=false,font=footnotesize]{subfig}
\fi

\usepackage{algorithm}
\usepackage[colorinlistoftodos]{todonotes}

\begin{document}
\title{Unit Circle Rectification of the MVDR Beamformer}
\author{Saurav~R.~Tuladhar,~\IEEEmembership{}
        John~R.~Buck,~\IEEEmembership{Senior Member,~IEEE}
\thanks{This work was supported by the Office of Naval Research grant N00014-12-1-0047 and N00014-15-1-2238}}
%
\onecolumn

\maketitle



\begin{abstract}
The sample matrix inversion (SMI) beamformer implements Capon's minimum variance distortionless (MVDR) beamforming using the sample covariance matrix (SCM). In a snapshot limited environment, the SCM is poorly conditioned resulting in a suboptimal performance from the SMI beamformer. Imposing structural constraints on the SCM estimate to satisfy known theoretical properties of the ensemble MVDR beamformer  
mitigates the impact of limited snapshots on the SMI beamformer performance. Toeplitz rectification and bounding the norm of weight vector are common approaches for such constrains. This paper proposes the unit circle rectification technique which constraints the SMI beamformer to satisfy a property of the ensemble MVDR beamformer: for narrowband planewave beamforming on a uniform linear array, the zeros of the MVDR weight array polynomial must fall on the unit circle. Numerical simulations show that the resulting unit circle MVDR (UC MVDR) beamformer frequently improves the suppression of both discrete interferers and white background noise compared to the classic SMI beamformer. Moreover, the UC MVDR beamformer is shown to suppress discrete interferers better than the MVDR beamformer diagonally loaded to maximize the SINR.  
\end{abstract}

\begin{IEEEkeywords}
adaptive beamformer, MVDR, Capon beamformer, array polynomial, unit circle 
\end{IEEEkeywords}


\IEEEpeerreviewmaketitle
\input{mycommands}

\input{introduction}
\input{background}

\input{arraypoly}
\input{ucmvdr}
\input{results}
\input{discussion}
\input{conclusion}


%

\appendices
\input{appendix}
\ifCLASSOPTIONcaptionsoff
  \newpage
\fi


\bibliographystyle{./refs/IEEEtran}
\bibliography{./refs/IEEEfull,./refs/myrefs}

\vfill


\end{document}

%% file: mycommands.tex
\newcommand{\Cov}{\boldsymbol{\Sigma}}
\newcommand{\cov}{\sigma}
\newcommand{\Covdmr}{\Cov_{\rm DMR}}
\newcommand{\eval}{\gamma}
\newcommand{\Eval}{\boldsymbol{\Gamma}}
\newcommand{\evec}{\boldsymbol{\xi}}
\newcommand{\Evec}{\boldsymbol{\Xi}}
\newcommand{\sampCov}{{\bf S}}
\newcommand{\sampcov}{s}
\newcommand{\sampCovdmr}{\sampCov_{\rm DMR}}
\newcommand{\sampeval}{g}
\newcommand{\sampEval}{{\bf G}}
\newcommand{\sampevec}{{\bf e}}
\newcommand{\sampEvec}{{\bf E}}
\newcommand{\herm}{^{\rm H}}
\newcommand{\trans}{^{\rm T}}
\newcommand{\rep}{{\bf v}}        
\newcommand{\repmat}{{\bf V}}     
\newcommand{\sigamp}{b}
\newcommand{\wmvdr}{{\bf w}_{\rm MVDR}}
\newcommand{\wdmr}{{\bf w}_{\rm DMR}}
\newcommand{\dl}{\delta}
\newcommand{\dlmin}{\dl_{min}}
\newcommand{\wconv}{{\bf w}_{\rm conv}}
\newcommand{\nulldepth}{\mbox{ND}}
\newcommand{\replook}{\rep_0}
\newcommand{\repint}{\rep_I}
\newcommand{\tsp}{^{\rm T}}
\newcommand{\inv}{^{-1}}
\newcommand{\limrmt}{\lim_{RMT}}
\newcommand{\eig}{\operatorname{eig}}
\newcommand{\diag}{\operatorname{diag}}
\newcommand{\sign}{{\operatorname{sgn}}}

\newcommand{\ulook}{u_0}
\newcommand{\uinter}{u_I}
\newcommand{\beampatu}{B(u)}
\newcommand{\beampat}[1]{B(#1)}
\newcommand{\wcbf}{{\bf w}_{\rm CBF}}
\newcommand{\wsmi}{{\bf w}_{\rm SMI}}
\newcommand{\wuc}{{\bf w}_{\rm UC}}
\newcommand{\wt}{{\bf w}}
\newcommand{\wtKuc}{\tilde{\wt}_{\rm Kuc}}
\newcommand{\wtK}{\wt_{\rm K}}
\newcommand{\wtdz}{\wt_{\rm DZ}}
\newcommand{\wthat}{\hat{\wt}}
\newcommand{\beampolyz}[1]{{P}_{#1}(z)}
\newcommand{\mvdrpoly}[1]{{P}_M(#1)}
\newcommand{\cbfpoly}{P_C(z)}
\newcommand{\smipoly}{P_S}
\newcommand{\ucpoly}{P_{UC}}
\newcommand{\datavec}{{\bf x}}
\newcommand{\dataMat}[1]{\mathbf{X}_{#1}}
\newcommand{\dataSubMat}[1]{\mathbf{Y}_{#1}}
\newcommand{\dataMatK}{\mathbf{X}_{\rm K}}
\newcommand{\noisevec}{{\bf n}}
\newcommand{\zvec}{\boldsymbol{z}}
\newcommand{\ztrans}{\mathcal{Z}}
\newcommand{\eye}{{\bf I}}     
\newcommand{\srt}[1]{\textbf{SRT:#1}}
\newcommand{\uc}{\mathcal{U}}
\newcommand{\cnormal}[2]{\mathcal{CN}(#1  #2)}
\newcommand{\sampz}{\xi}
\newcommand{\ensz}{\zeta}
\newcommand{\ucz}{\hat{\xi}}
\newcommand{\zinter}{\ensz{}_I}
\newcommand{\pout}{P_{\rm out}}
\newcommand{\poutmvdr}{{\rm P}_{\rm MVDR}}
\newcommand{\notchdepth}{\text{ND}}
\newcommand{\powspec}{S(u)}
\newcommand{\antieye}{\mathbf{J}}
\newcommand{\sgain}{G}
\newcommand{\dataMatsa}{\mathbf{X}_{\rm SA}}
\newcommand{\bigO}[1]{\mathcal{O}(#1)}
\newcommand{\taperMat}{\mathbf{T}}
\newcommand{\expo}[1]{e^{#1}}

\newcommand{\scov}{\ensuremath{\hat{\bf R}}}
\newcommand{\ncov}{\ensuremath{\bf{R}_n}}
\newcommand{\nicov}{\ensuremath{\bf{R}_{i+n}}}
\newcommand{\Knn}{\ensuremath{\mathbf{R}_n}}
\newcommand{\sKnn}{\ensuremath{\hat{\mathbf{R}}_n}}
\newcommand{\spow}{\ensuremath{\sigma_s^2}}
\newcommand{\tpl}{\frac{2\pi}{\lambda}}

\newcommand{\noisepow}{\sigma^2_W}
\newcommand{\interfpow}{\sigma_I^2}
\newcommand{\avgCov}{\Cov_A}
\newcommand{\cmtW}{W}
\newcommand{\resW}{R_{\Delta u}}

\newcommand{\sigvec}{\mathbf{x}}  
\newcommand{\sigvecW}{\mathbf{X}} 
\newcommand{\sig}{x}
\newcommand{\sigW}{X}  
\newcommand{\sigWvec}{\mathbf{X}}
\newcommand{\pos}{\mathbf{p}} 
\newcommand{\hresp}{h}
\newcommand{\hrespW}{H}
\newcommand{\hrespWvec}{\mathbf{H}}
\newcommand{\yt}{y}  
\newcommand{\yw}{Y}  
\newcommand{\wavenum}{\mathbf{k}}
\newcommand{\pnoise}{\mathcal{P}_{\rm N}}
\newcommand{\pinter}{\mathcal{P}_{\rm I}}
\newcommand{\half}{\frac{1}{2}}
\newcommand{\tr}[1]{\rm tr(#1)}
\newcommand{\delay}{\tau}
\newcommand{\deltaT}{\Delta T}
\newcommand{\taperevec}{\boldsymbol{u}}
\newcommand{\modvec}{\boldsymbol{y}}  

\newcommand{\mbf}[1]{\mathbf{#1}}
\newcommand{\fig}{Fig.~}
\newcommand{\sect}{Sec.~}
\newcommand{\eqn}{Eq.~}
\newcommand{\nth}{^{\rm th}}
\newcommand{\norm}[1]{||#1||}

\newcommand{\wts}{\wt}
\newcommand{\x}{\datavec}
\newcommand{\X}{\dataMat{}}
\newcommand{\cbfz}{z}

\newcommand{\deltau}{\Delta u}
\newcommand{\inr}{\eta^2_I}
\newcommand{\wng}{\text{WNG}}
\newcommand{\cbfNtwozeropoly}{F_C}
\newcommand{\funudu}{f(u, \deltau)}
\newcommand{\funczero}[1]{f(#1, \deltau)}

\newcommand{\mvdroutput}{P_{\rm MVDR}}
\newcommand{\interfout}{\mathcal{P}_I}
\newcommand{\noiseout}{\mathcal{P}_W}
\newcommand{\interz}{\ensz_I}

\newcommand{\puc}{P_{\rm UC}}
\newcommand{\pone}{P_{1}}
\newcommand{\pdelta}{P_{\Delta}}


%% file: introduction.tex
\section{Introduction}
\label{sec:intro}
Adaptive beamformers in practical settings face the challenge of extracting enough information from a limited set of observations, or snapshots, to suppress discrete interferers while maintaining attenuation of background white noise. The motion of sources enforces a limit on the number of snapshots which can be coherently combined \cite{baggeroer1999passive}. Imposing structural constraints on the covariance matrix or array weights forces a beamformer to satisfy known theoretical properties of the ensemble covariance matrix (ECM), making the best use of the limited information available in snapshot poor scenarios. Common approaches for such constraints include Toeplitz rectification \cite{cadzow1987,barton1997,vallet2014,vallet2017performance,fuhrmann1991toeplitz}, and limiting the norm of the weight vector \cite{Cox1987robust}. Imposing an upper bound on the norm of the weight vector ensures a minimum amount of white noise gain while also providing robustness to mismatch in sensors' locations or gains \cite{Gilbert1955}. In the spirit of such constraints, this paper proposes a new rectification technique for mitigating the impact of limited snapshots on the performance of Capon's minimum variance distortionless response (MVDR) beamformer \cite{capon1969mvdr} - the unit circle rectification.

The unit circle rectification constrains the MVDR beamformer weights to satisfy a property first observed by Steihardt and Guerci: for a narrowband planewave MVDR beamformer on a uniform linear array (ULA), the zeros of the $z$-transform of the conjugated array weights must fall on the unit circle \cite{steinhardt2004stap}. The polynomial resulting from this $z$-transform is known as the \textit{array polynomial} \cite{Schelkunoff1943array}. This unit circle property observed by Steinhardt and Guerci holds for the case of the ensemble MVDR beamformer. However, the zeros of the array polynomial of the practical sample matrix inversion (SMI) beamformer generally do not fall on the unit circle. The unit circle rectification enforces consistency with the unit circle property by projecting the zeros of the SMI beamformer array polynomial back to the unit circle \cite{tuladhar2015ucmvdr}. The projection can be considered as an alternative method of conditioning the beamformer weights, similar in spirit but complementary to Toeplitz rectification \cite{cadzow1987,barton1997,vallet2014, vallet2017performance, fuhrmann1991toeplitz}. Enforcing this consistency moves the SMI solution closer to the ideal ensemble solution, improving the performance of the practical adaptive beamformer. 

This paper focuses on transferring intuition about the $z$-transform and the frequency response in time-domain linear systems to the array polynomial and beampattern in spatial processing, rather than rigorous performance proofs.  Monte Carlo simulations for small and large arrays for several snapshot regimes confirm the benefits of transferring this intuition across domains. The resulting unit circle MVDR (UC MVDR) beamformer is remarkable in that simulation results show that for the snapshot limited scenarios studied here, it frequently improves the suppression of both discrete interferers and white background noise relative to the classic SMI solution. 


The remainder of this paper is organized as follows:
Sec.~\ref{sec:bkgnd} reviews planewave beamforming and establishes the notation followed in the sequel. Sec.~\ref{sec:array-poly} develops the array polynomial representation for narrowband beamformers using ULAs. \sect{}\ref{sec:ucmvdr-algorithm}
presents the unit circle rectification technique used to derive the UC MVDR beamformer. Sec.~\ref{sec:ucbf-perf}
discusses the simulation results comparing the performance of the UC
MVDR beamformer against the SMI beamformer and the diagonal loaded MVDR beamformer. Sec.~\ref{sec:discussion} discusses the limitations of the proposed unit circle rectification technique and the potential directions for future enhancement.  Sec.~\ref{sec:conclusion} presents the concluding remarks.


%% file: background.tex
\section{Planewave beamforming}
\label{sec:bkgnd}
This section presents a brief review on planewave beamforming
to establish the notation used in this paper. A comprehensive
discussion on planewave beamforming and related topics can be found in
the classical texts \cite{vtree2002oap, johnson1992array}.

The narrowband planewave data measured with an $N$ sensor ULA is represented as an $N \times 1$ vector of complex phasors $\datavec$ or a snapshot, commonly modeled as, 

\begin{equation}
  \label{eq:array-data} \datavec = \sum\limits_{i=1}^D a_i\rep_i +
\noisevec,
\end{equation}
where $D$ is the number of planewave signals, $a_i$ is $i^{th}$ signal
amplitude, $\rep_i$ is the planewave array manifold vector for the
$i^{th}$ signal, and $\noisevec$ is the additive noise vector. The
amplitudes are modeled as a zero mean complex circular Gaussian random
variable, i.e., $a_i \sim \cnormal{0, \cov_i^2}$ and the background
noise is assumed to be spatially white with complex circular Gaussian
distribution, i.e.,
$\noisevec \sim \cnormal{\mathbf{0}, \cov_w^2\eye}$. The array
manifold vector is a complex exponential vector

\begin{equation*}
\label{eq:array-manifold}
  \rep_i = [1,~e^{-j{(2\pi u_i/\lambda)} d},~ e^{-j{(2\pi u_i/\lambda)}2d},
\ldots, e^{-j{(2\pi u_i/\lambda)}(N-1) d}]^\text{T},
\end{equation*}
where $u_i = \cos(\theta_i)$ and $\theta_i$ is the $i^{th}$ signal
direction, $\lambda$ is the wavelength, $d$ is the inter-sensor
spacing on the ULA and $[\cdot]^T$ denotes transpose. In the sequel,
the signal direction will be represented in terms of the directional
cosine $u$. Assuming the $D$ signals in \eqref{eq:array-data} are
uncorrelated, the data ECM is

\begin{align}
  \label{eq:ecm} \Cov =& E[\datavec\datavec\herm] = \sum\limits_{i =
1}^D\cov_i^2\rep_i\rep_i\herm + \cov_w^2\eye,
\end{align}
where $\cov_i^2$ is $i^{th}$ signal power and $\cov_w^2$ is the sensor
level noise power. The ECM quantifies the statistical relation between the data measured at each pair of sensors.

Beamformers spatially filter the array data by passing the signal from
select look directions while rejecting noise and interfering signals.
The beamformer output ($y$) is obtained as the weighted sum of
the sensor data, i.e. $y = \wt\herm\datavec$, where $\wt$ is the $N\times 1$ complex
array weight vector. The conventional beamformer (CBF) is the basic 
beamformer whose weights shift signals measured at each sensor such that 
planewave signals from the look direction $\ulook$ align in time and 
combine constructively while signals from other directions combine 
destructively and are suppressed. A CBF steered to the look direction 
$\ulook = \cos(\theta_0)$ has a weight vector equal to the array 
manifold vector for the look direction normalized for unit gain, i.e., $\wcbf = \rep_0/N$.

 The choice of weights determines the
beamformer's beampattern. The beampattern $\beampatu$ defines the
complex gain due to the beamformer for a unit amplitude planewave from
direction $u = \cos(\theta)$, i.e.,

\begin{equation}
  \label{eq:beampatu} 
  \beampatu = \wt^H\rep(u) = \sum_{n=0}^{N-1}w_n^*\left(e^{-j\frac{2\pi u}{\lambda} d}\right)^n
\end{equation}
where $(\cdot)^*$ denotes conjugate and $-1 \leq u \leq 1$ is the
directional cosine. The beampattern is analogous to the frequency response of a discrete-time (DT) linear system. The beampattern magnitude $|\beampatu|$ is characterized by a mainlobe in the look direction ($u_0$) and the multiple sidelobes outside the mainlobe. The beampattern magnitude squared in the interferer direction is the notch depth (ND), i.e., $\notchdepth = |\beampat{\uinter}|^2$, where $\uinter$ is the interferer direction. The ND quantifies the interferer attenuation due to the beamforming.

White noise gain quantifies the improvement in SNR due to
the beamformer suppressing the spatially-white background noise. Assuming
unity gain in the look direction i.e., $\beampat{u_0} = 1$, white noise
gain (WNG) is given by $\text{WNG} = ||\wt||^{-2}$ where
$||\cdot||$ denotes the Euclidean norm \cite{vtree2002oap}. WNG is
also a metric for beamformer robustness against mismatch
\cite{Gilbert1955}. For a given ULA, the CBF has the maximum WNG which
is equal to the number of array elements $N$ \cite{Cox1987robust}.



\subsection{MVDR beamformer}
\label{sec:mvdr-beamformer}
Capon proposed the MVDR beamformer as an optimum adaptive beamformer (ABF) \cite{capon1969mvdr}. The MVDR beamformer minimizes the output variance $E(|y|^2) = \wt\herm\Cov\wt $ while maintaining unity gain in the look direction with the array manifold vector $\replook$. The MVDR weight vector $\wmvdr$ is derived as the solution to the constrained optimization problem

\begin{equation}
  \label{eq:mvdr-const-prob}
  \underset{\wt}\min \quad f(\wt) = \wt\herm\Cov\wt \quad \text{s.t.} \quad \wt\herm\replook = 1,
\end{equation}
where $\Cov$ is the interferer-plus-noise ECM and $\replook$ is the
array manifold vector for the look direction
$\ulook = \cos(\theta_0)$. The optimal solution is

\begin{equation}
  \label{eq:mvdr-wt} 
\wmvdr =
\frac{\Cov\inv\replook}{\replook\herm\Cov\inv\replook}.
\end{equation}

 Computing the MVDR weights in \eqref{eq:mvdr-wt} requires knowledge of the ECM but in practical applications the ECM is unknown. Consequently,
the MVDR beamformer is approximated by the SMI beamformer
by replacing the ECM $\Cov$ in \eqref{eq:mvdr-wt} with the SCM,

\[ 
\sampCov = \frac{1}{L}\sum\limits_{l=1}^{L}\datavec_l\datavec_l\herm
\]
where the $L$ is the number of data snapshots and $\datavec_l$ is the
$l^{th}$ data snapshot vector defined in \eqref{eq:array-data}. The SMI beamformer is the simplest practical implementation of the MVDR beamformer.

The SMI beamformer relies on the availability of a large number of
snapshots ($L$) such that $L \gg N$, where $N$ is the number of sensors. The SCM computed from sufficiently large $L$ gives an accurate estimate of the
ECM \cite{boroson1980sample}. Reed et al.\ show that at least two
snapshots per sensor, i.e., $L \geq 2N$ is required to ensure that the
expected output SINR loss due to the use of the SCM instead of the ECM
is $3$ dB or less \cite{reed1974rapid}. In many beamforming scenarios,
physical non-stationarities in the environment or source locations
preclude averaging large numbers of snapshots to form the SCM. The use
of long arrays and the presence of fast moving sources severely limit
the available number of snapshots. In many passive sonar applications
it is common to have only limited snapshots ($L \approx N$) or even
insufficient snapshots ($L < N$) available and even two snapshots per
sensor ($L = 2N$) is considered a snapshot rich case
\cite{baggeroer1999passive, cox2000mrabf}. When the number of
snapshots are limited ($L \approx N$) the SCM is ill-conditioned and
the SCM inversion is numerically unstable. Inadequate estimation of
the SCM results in high sidelobes and distorted mainlobe in the
beampattern and subsequent degradation in interferer and white noise
attenuation \cite{Carlson1988scm, richmond2000mvdr}. Moreover, in the
snapshot deficient case $L < N$, the SCM is rank deficient and the SCM
inversion is not possible. This paper focuses on the limited snapshot scenarios with $L \leq 2N$.


A common approach to address the limited snapshot scenario is to apply
diagonal loading (DL) to the SCM to get $\sampCov_\dl = \sampCov + \dl\eye{}$, where $\dl$ is the DL factor
\cite{vtree2002oap}. The DL MVDR beamformer weights are computed by
replacing the ECM in \eqref{eq:mvdr-wt} with the DL SCM $\sampCov_\dl$. DL makes the SCM inversion numerically stable, provides better sidelobe
control and improves the beamformer WNG while introducing bias
\cite{cox2002adaptive, mestre2005diagonal, nadakuditi2005bias}. 


However, choosing the best loading factor to combat the impact of limited snapshots in a practical scenario remains a challenging problem. Several ad-hoc approaches have been proposed for choosing the appropriate DL factor: $\dl = 10\sigma^2$ where $\sigma^2$ is the noise power \cite{gershman2003robust}, $\dl = -\hat{\lambda}_{min} + \sqrt{(\hat{\lambda}_D - \hat{\lambda}_{min})(\hat{\lambda}_{D+1} - \hat{\lambda}_{min})}$ where $\hat{\lambda}_{max} = \hat{\lambda}_1 \geq \ldots \geq \hat{\lambda}_N = \hat{\lambda}_{min}$ are the sample eigenvalues and $\rm D$ is dimension of the interferer subspace. These methods require knowledge of the noise power level and interferer subspace dimension which are unknown in practice. Other ad-hoc considerations for the choice of diagonal loading have been proposed as summarized in \cite[Sec:III]{mestre2006finite}.

In contrast, Mestre and Lagunas systematically optimized the DL factor to maximize the SINR to derive 
a random matrix theory based optimal DL factor estimator for snapshot limited scenarios \cite{mestre2005diagonal}. The
authors derive an expression relating the asymptotic output SINR to
the DL factor $\dl$ and the ratio of snapshots per sensor. The optimal
DL factor is the solution that maximizes the output SINR. However, the
procedure to search for the optimal DL factor has a significantly
higher computational complexity compared to ad-hoc
procedures.



%% file: arraypoly.tex
\section{Array Polynomials}
\label{sec:array-poly}
The array polynomial is the $z$-transform of the conjugated array weights of a narrowband beamformer for a ULA \cite{Schelkunoff1943array, vtree2002oap, Steinberg1976}. The array polynomial is analogous to the system function representation of a DT linear system obtained by taking the $z$-transform of the system impulse response. As with DT linear systems, beamformers also have a pole-zero representation in the complex plane. Continuing the analogy, the beampattern is obtained by evaluating the array polynomial along the unit circle $\lbrace z \in \mathbb{C}, |z| = 1\rbrace$. 

The beampattern of a narrowband beamformer using a ULA can be
represented as a complex polynomial
\cite{Schelkunoff1943array}\cite{Steinberg1976}. Letting
$d = \lambda/2$ and $z = e^{j\pi u}$ in \eqref{eq:beampatu} yields the
array polynomial

\begin{equation}
  \label{eq:beampat-poly}
  \beampolyz{} = \sum\limits_{n=0}^{N-1} w^*_n z^{-n} = \ztrans(\wt\herm).
\end{equation}
where $\ztrans()$ is the z-transform operator and $*$ denotes the complex 
conjugate operator. $\beampolyz{}$ is an $N-1$ degree polynomial in complex 
variable $z$ with the conjugated complex beamformer weights ($w^*_n$) as
coefficients. \eqn\eqref{eq:beampat-poly} is the
$z$-transform of the conjugated beamformer weights
\cite[Chap.~3]{Oppenheim1989}. The array polynomial representation
maps the bearing variable $u$ into the complex plane. The phase of the
complex variable ($z$) is related to the cosine bearing variable $u$
as $\operatorname{arg}(z) = \pi u$. In the complex plane, the array
polynomial has $N - 1$ zeros and the corresponding $N - 1$ poles are
all at the origin. Evaluating \eqref{eq:beampat-poly} on the unit
circle $\lbrace z \in \mathbb{C}, |z| = 1\rbrace$ yields the
beampattern. The zeros of the array polynomial correspond to the beampattern notches and when the zeros fall on the unit circle they result in perfect notches or \emph{nulls} in the beampattern.

A CBF using an $N$ sensor ULA and steered to the broadside ($\ulook = 0$) look direction has a weight vector $\wcbf = {\bf 1}/N$ where ${\bf 1}$ is a vector of $N$ ones. Applying the array polynomial approach to the CBF finds the classic finite geometric series result from linear system theory, possibly shifted in angle based on the look direction. The $z$-transform of $\wcbf$ gives the CBF polynomial
\begin{align}
  \label{eq:cbf-poly}
  \cbfpoly = \frac{1}{N}\sum\limits_{n=0}^{N-1}z^{-n} 
  = \frac{1}{N}\left[\frac{z^N - 1}{z^{N-1}(z - 1)}\right].
\end{align}
When $\cbfpoly$ is evaluated on the unit circle, the finite geometric series can be manipulated into the well-known discrete sinc function, with a main lobe in the look direction \cite[Chap.~3]{Oppenheim1989}. The CBF polynomial zeros are the roots of the numerator in \eqref{eq:cbf-poly}, which are the $N$ roots of unity. These roots give equally spaced zeros on the unit circle 
except for the root at $z = 1$ which is canceled by a corresponding pole. Hence, a CBF using an $N$ sensor ULA has  $N - 1$ zeros confined onto the unit circle and these zeros produce the nulls in the CBF beampattern \cite{vtree2002oap}.




\subsection{MVDR array polynomial}
\label{sec:mvdr-poly}
\begin{figure}[!t]
  \centering
  \subfloat[]{\includegraphics[width=3in]{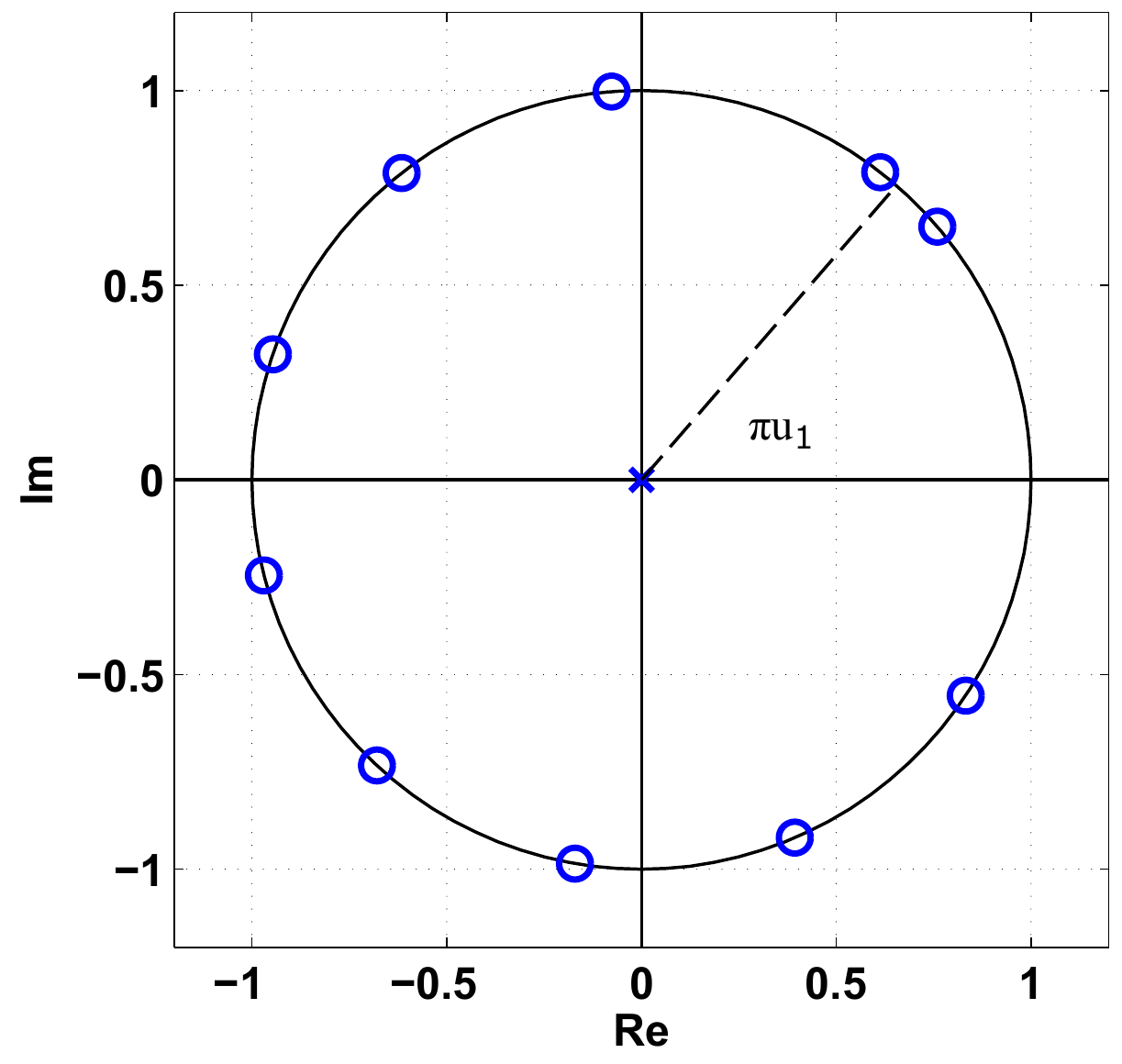}
    \label{fig:mvdr-pzplot}} \hfill
  \subfloat[]{\includegraphics[width=3in]{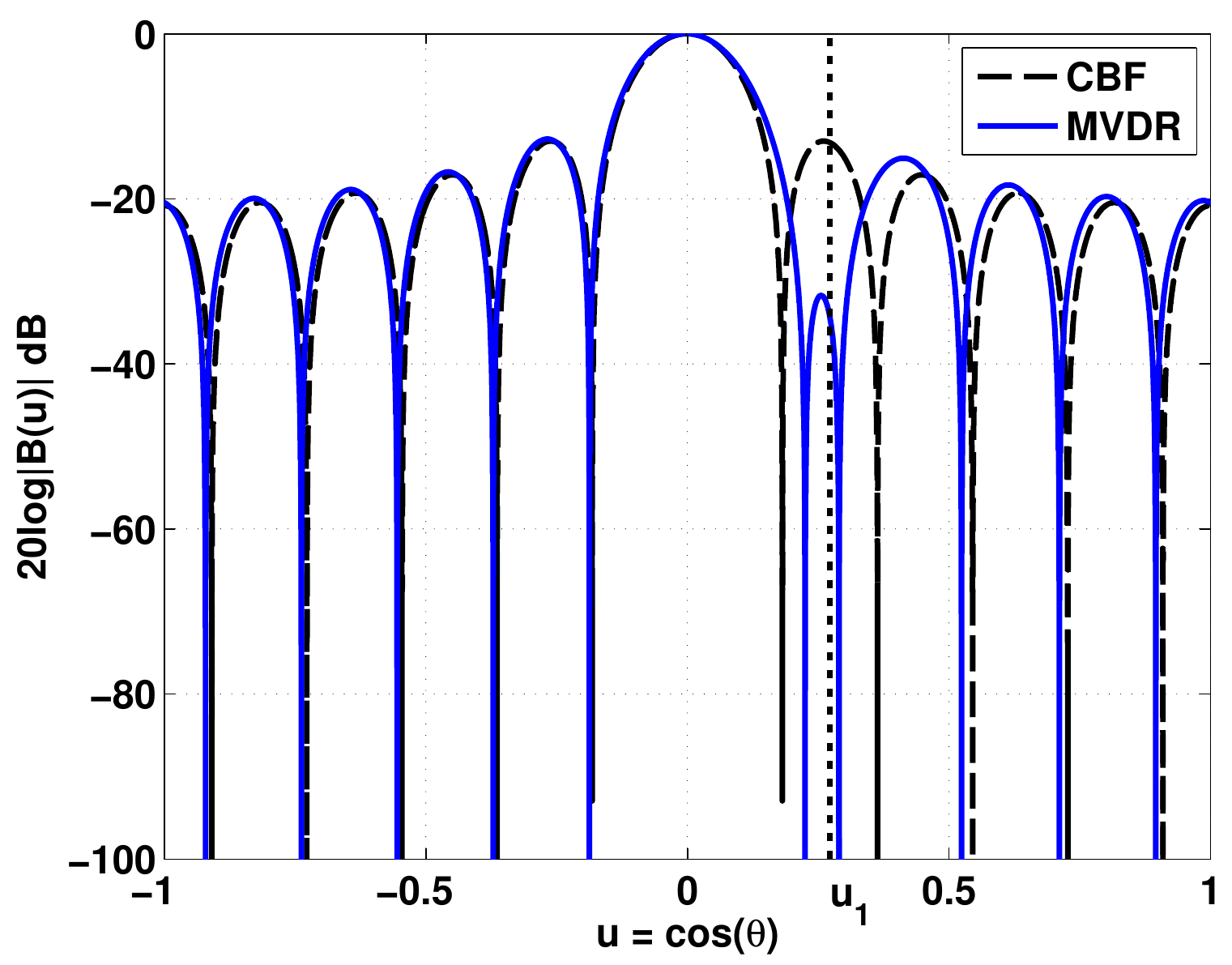}
  \label{fig:cbf-mvdr-bpplot}}
\caption{MVDR beamformer \protect\subref{fig:mvdr-pzplot} zero locations
  and \protect\subref{fig:cbf-mvdr-bpplot} beampattern for an example case
  using $N = 11$ sensor ULA and a single interferer at $u_1 = 3/N$
  indicated by vertical dashed line in
  \protect\subref{fig:cbf-mvdr-bpplot}. In
  \protect\subref{fig:mvdr-pzplot} the dashed
  line at $\pi u_1$ indicates the angle in the complex plane
  corresponding to the interferer.}
  \label{fig:mvdr-plots}
\end{figure}

Following the definition in \eqref{eq:beampat-poly}, the ensemble MVDR beamformer 
array polynomial is $\mvdrpoly{z} = \ztrans(\wmvdr\herm)$. Factoring
the polynomial,
\[
\mvdrpoly{z}  =  \Gamma \prod\limits_{n=1}^{N-1}(1 - \ensz_n z\inv),
\]
where $\Gamma$ is a scaling term and $\ensz_n$ is the $n^{th}$
zero of the ensemble MVDR beamformer array polynomial (or "ensemble zero" for brevity in the sequel). \figurename{}~\ref{fig:mvdr-plots} shows the ensemble
zeros and beampattern for an example case of the MVDR beamformer using an
$N = 11$ sensor ULA. A single interferer is present at $\uinter = 3/N$
and the interferer-to-noise power ratio (INR) is 10 dB. The dashed radial
line in \figurename{}~\ref{fig:mvdr-pzplot} indicates the angle
$\pi\uinter$ corresponding to the interferer direction $\uinter$
denoted by the vertical dashed line in
\figurename{}~\ref{fig:cbf-mvdr-bpplot}. All $N-1$ ensemble zeros in
\figurename{}~\ref{fig:mvdr-pzplot} are on the unit circle and these
zeros correspond to the beampattern nulls in
\figurename{}~\ref{fig:cbf-mvdr-bpplot}. However, the beampattern
nulls are not necessarily in the interferer direction. The interferer
ND depends on the INR level. The objective function in \eqref{eq:mvdr-const-prob} requires that the MVDR beamformer minimizes the total output power. As the INR changes, the MVDR beamformer adapts the ND and the WNG to reduce the total output power. This behavior manifests in the form of MVDR polynomial zeros shifting along the unit circle as the INR changes. Numerical experiments show that the MVDR beamformer controls the interferer suppression by placing a beampattern notch such that the interferer falls on the shoulder of the notch. As INR increases, the zeros shift along the unit circle towards the interferer direction yielding a deeper notch. As INR decreases, the zeros shift away from the interferer direction yielding a shallower notch. In fact the MVDR ensemble zeros are always located on the unit circle for planewave beamforming using ULAs. Steinhardt and Guerci \cite{steinhardt2004stap} first proved this unit circle property for the ensemble MVDR beamformer, though their result does not seem to be widely known. Appendix~\ref{sec:apdx-mvdr-zeros} outlines their proof of the unit circle property.

\begin{figure}[!t]
  \centering
  \includegraphics[width=3in]{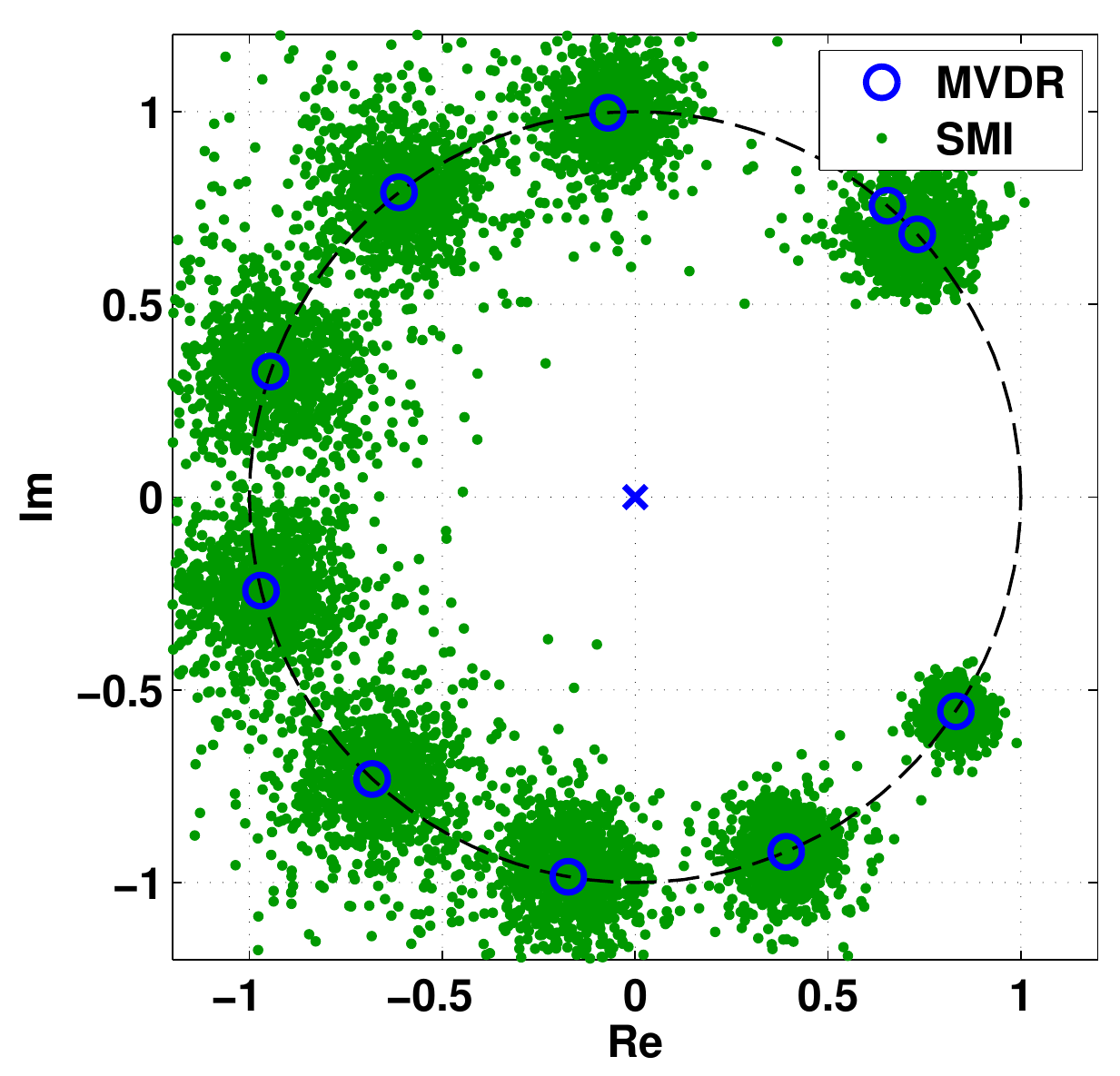}
  \caption{SMI MVDR zeros from 1000 Monte Carlo trials for N = 11, L =
    10N and INR = 40 dB. The number of snapshots is impractically
    large for many passive sonar situations, but makes clear the
    clustering of the SMI zeroes around the ensemble zeroes.}
  \label{fig:smi-mvdr-pzplot}
\end{figure}

In practice, each realization of the SMI MVDR beamformer has an array
polynomial representation defined as $\smipoly = \ztrans(\wsmi\herm)$.
The zeros of the SMI MVDR weights array polynomial (or "sample zeros" in the sequel)  are randomly perturbed from the ensemble
zero locations on the complex
plane. \figurename{}~\ref{fig:smi-mvdr-pzplot} is a composite of
sample zeros (green markers) obtained from 1000 independent
realizations of the SMI MVDR beamformer. It considers an example case
of a $N = 11$ sensor ULA using $L = 10N$ snapshots to compute the SCM and a
single interferer present $\uinter = 3/N$ with INR = 40 dB. The number of snapshots 
is impractically large for many passive sonar situations, but is chosen to create a 
clearer clustering of the sample zeros around the ensemble zeros. Examining 
\figurename{}~\ref{fig:smi-mvdr-pzplot} shows the sample zeros clustering around the 
ensemble zero locations $\ensz_n$ while not necessarily falling on the unit circle. 
The SMI MVDR beampattern converges in probability to the ensemble beampattern as the 
number of snapshots $L$ increases \cite{richmond2000nulling}. Thus the sample zeros 
also converge to ensemble zero locations as the number of snapshot increases.

Continuing the analogy with DT LTI systems, the sample zeros that fall
away from the unit circle correspond to shallow notches instead of
nulls in SMI MVDR beampattern. Further, any zeros that fall closer to
the origin or far outside the unit circle have negligible contribution
to beampattern \cite[Chap.~5]{Oppenheim1989}. Hence the SMI MVDR beamformer suffers
from beampattern distortion resulting in loss of interferer
suppression and WNG. The following section describes how the SMI MVDR
beamformer can be modified to improve interferer suppression by moving
the sample zeros to the unit circle.

%% file: ucmvdr.tex
\section{Unit circle rectification}
\label{sec:ucmvdr-algorithm}
The unit circle rectification algorithm projects the sample zeros radially to the
unit circle, enforcing the unit circle property of the ensemble MVDR polynomial 
zeros. Applying the unit circle rectification to the SMI beamformer produces 
the UC MVDR beamformer. The zeros on the unit circle guarantee nulls in the UC MVDR 
beampattern in the directions corresponding to the sample zeros. The rest of the 
section describes the unit circle rectification to derive the UC MVDR 
beamformer assuming the beamformer is steered towards the broadside ($\ulook = 0$). 
The algorithm extends naturally to the case of a different look direction ($\ulook = 
u_L$) by changing the array manifold vector used for the SMI beamformer, and then shifting the mainlobe exclusion region for zeros described below.

\begin{figure}[tp]
\centering
\includegraphics[width=3in]{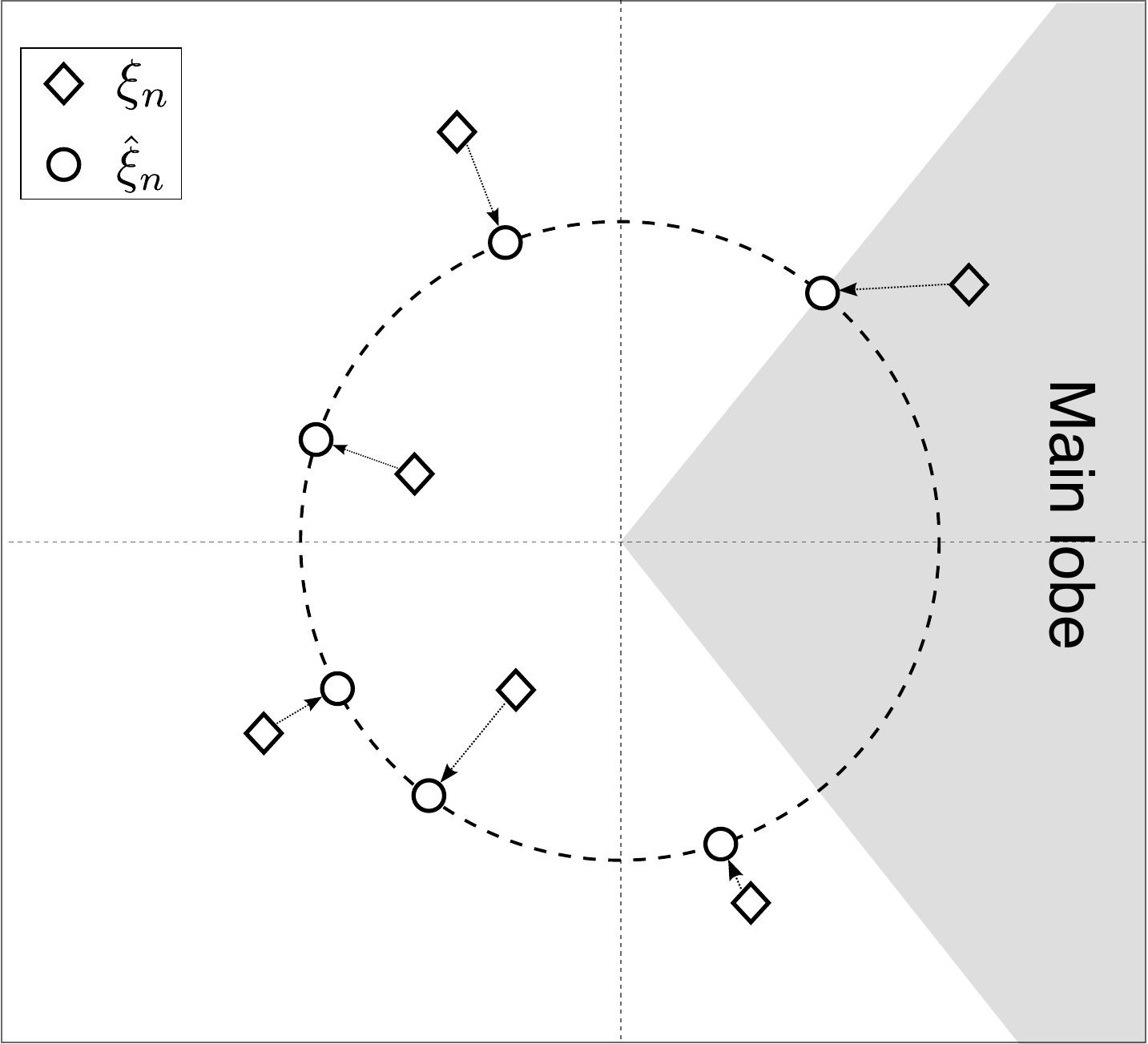}  
\caption[Schematic shows the unit circle projection
technique.]{Schematic shows the unit circle projection technique. The
  diamond markers denote the sample zeros ($\sampz_n$) and the circle
  markers denote the unit circle zeros ($\ucz$) obtained by radially
  projecting the sample zeros. The sample zeros within the main-lobe
  region ($-2/N \leq u \leq 2/N$) are moved to the closest CBF
  first-null location ($u_{\text{null}} = \pm 2/N$) on the unit
  circle.}
\label{fig:ucmvdr-cartoon}
\end{figure}

Algorithm~\ref{alg:ucmvdr} outlines the UC MVDR beamformer
implementation. The algorithm begins from the SMI MVDR weights
$\wsmi$. The $z$-transform of the conjugated weights gives the SMI
MVDR polynomial $\smipoly(z) = \ztrans{(\wsmi\herm)}$. Factoring the SMI
MVDR polynomial,
\[
\smipoly(z) = G \prod\limits_{n=1}^{N-1}(1 - \sampz_n z\inv),  
\]
where $G$ is the gain required to ensure unity gain in the look
direction ($\ulook = 0$), i.e., $\smipoly(\expo{j\pi\ulook}) = 1$ and
$\sampz_n = r_ne^{j\omega_n}$ is the $n^{th}$ SMI MVDR polynomial
zero. As previously discussed in \sect{}\ref{sec:mvdr-poly}, the
sample zeros are not necessarily on the unit circle and hence the
magnitudes $|\sampz_n| = r_n$ of the roots are generally not
unity. The next step is to radially project the SMI MVDR polynomial
zeros $\sampz_n$ to the unit
circle, which is essentially the unit circle rectification. \figurename{}~\ref{fig:ucmvdr-cartoon} illustrates the
projection of the sample zeros (diamond markers) to the unit circle
zeros (circle markers). Projection yields a set of unit circle zeros
$\ucz_n = e^{j\omega_n}$. An exception
occurs when the sample zeros fall within the CBF main-lobe region in
the complex plane
i.e. $|\operatorname{arg}(\sampz_n) - \pi\ulook|<2\pi/N$. Projecting such zeros
radially to the unit circle results in nulls inside the main-lobe of
the UC MVDR ABF. A null inside the main-lobe results in undesired
 main-lobe distortion and drastic loss in WNG
\cite[Sec.~6.3.1]{vtree2002oap}. Rather than radially projecting to
the unit circle, such zeros are moved to the closest CBF first-null
location on the unit circle such that
$\operatorname{arg}(\ucz_n) = \sign(\omega_n){2\pi/N}$ where
$\sign(\cdot)$ is the sign function. This exclusion strategy is functionally similar to other main-lobe protection schemes for adaptive beamformers use to avoid deep notches in the main lobe too close to the look direction, e.g., \cite{cox1997robust,cox1998matchfield, kogon2002robust}. The UC MVDR approach differs in manipulating the main-lobe nulls in terms of the locations of the array polynomial roots rather than the inner product of the look direction replica vector with the SCM eigenvectors.

The projected unit circle zeros $\ucz_n$ are used to synthesize a unit
circle polynomial

\begin{align}
\label{eq:uc-poly-rational}
\ucpoly(z) =& \prod\limits_{n=1}^{N-1}\frac{(1 - \ucz_n z\inv)}{(1 - \ucz_n)}.  
\end{align}
Rewriting the polynomial in terms of the coefficients

\begin{align}
\label{eq:uc-poly}
\ucpoly(z) =& \sum\limits_{n=0}^{N-1} c_n^* z\inv.
\end{align}
Comparing \eqref{eq:uc-poly} to the definition of the array polynomial
\eqref{eq:beampat-poly}, the coefficients $c_n$s can be viewed as
beamformer weights. Thus, the UC MVDR ABF weight vector is defined as
$\wuc = [c_1, c_2, \ldots, c_N]\trans$. The denominator of $(1 - \ucz_n)$ in \eqref{eq:uc-poly-rational} guarantees that the coefficients $c_n$ sum to 1, and thus satisfy the unity gain constraint on the broadside look direction ($\ulook = 0$) we are assuming here. Evaluating $\ucpoly(z)$ on the
unit circle produces the UC MVDR ABF beampattern with $N-1$ nulls in
the directions corresponding to $\ucz_n$'s and a unity gain in the look
direction, i.e., $\ucpoly(\expo{j\pi\ulook}) = 1$.

\begin{figure}[tp]
\centering
\subfloat[Zero locations]
{\includegraphics[width=3.0in]{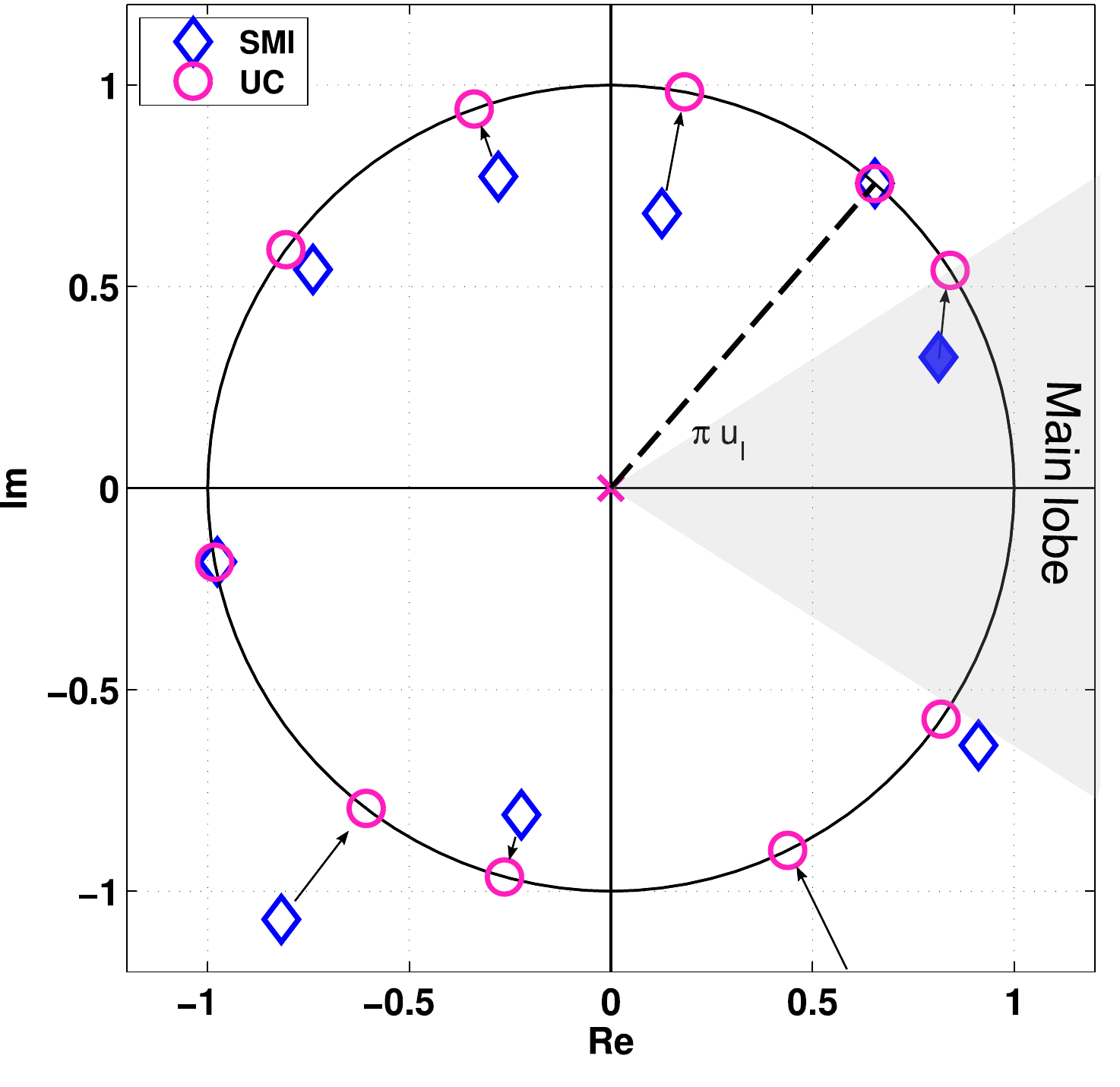}  \label{fig:smi-ucmvdr-pzplot}}

\subfloat[Log-magnitude beampattern]{\includegraphics[width=3.2in]{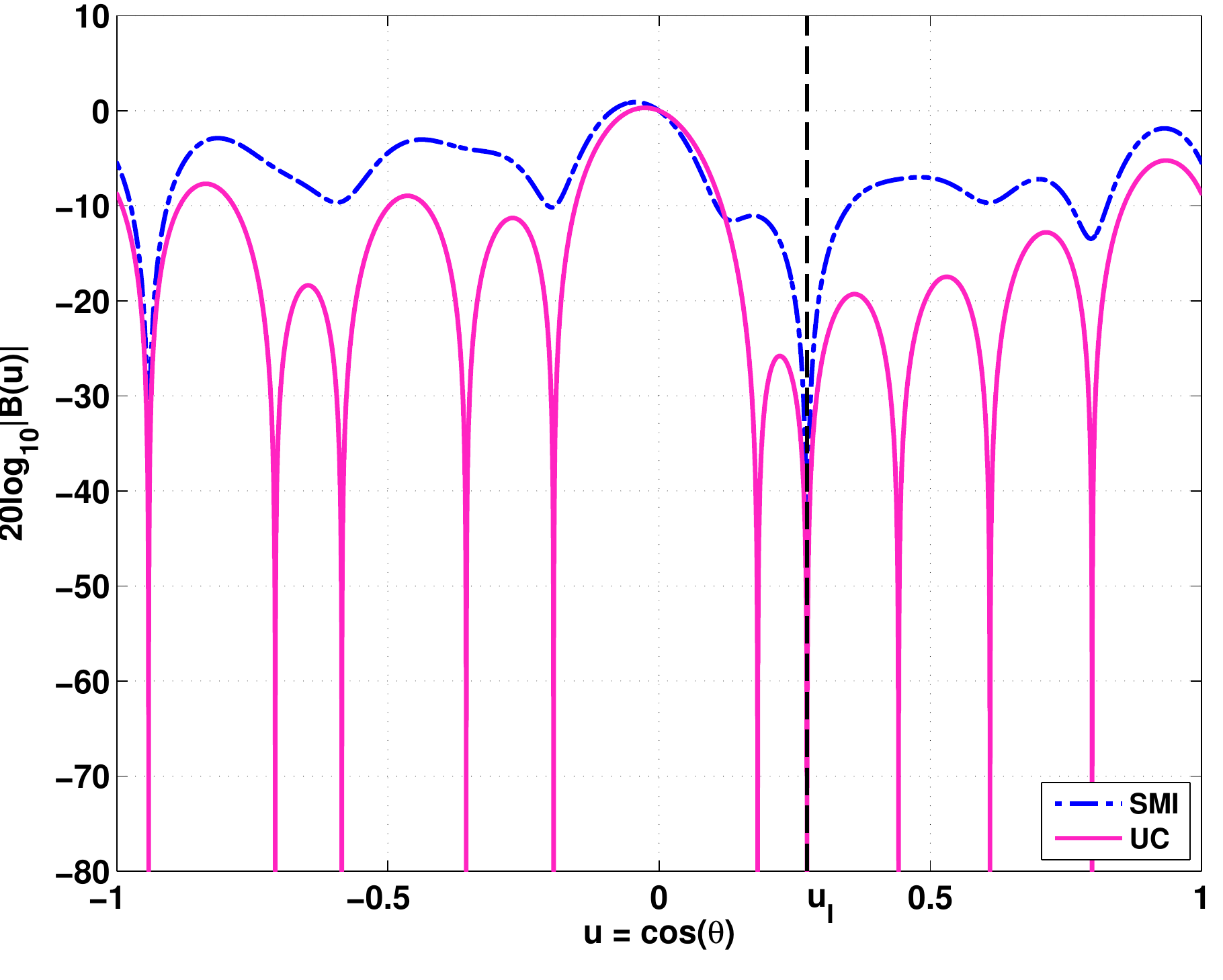}  \label{fig:smi-ucmvdr-bpplot}}
\caption[Zero locations and log-magnitude beampatterns of a
representative example of the SMI MVDR and the UC MVDR ABF.] {Zero locations
  and log-magnitude beampatterns of a representative example of SMI
  MVDR (blue) and UC MVDR ABF (magenta) using $N = 11$ sensor ULA for
  $L = 12$ snapshots and look direction $\ulook = 0$. The unit circle
  zeros of the UC MVDR ABF produce nulls and lowers sidelobes compared
  to the SMI MVDR ABF beampattern.}
\label{fig:smi-ucmvdr-plots}
\end{figure}

\figurename{} \ref{fig:smi-ucmvdr-plots} shows a representative
example comparing the UC MVDR and the
SMI MVDR ABF using an $N = 11$ sensor ULA and $L = 12$ snapshots. Both
ABFs are steered to broadside ($\ulook = 0$) look direction and a
single interferer is present at $\uinter = \cos(\theta_I) = 3/N$ with power $40$~dB above the background white noise. In
\figurename{}~\ref{fig:smi-ucmvdr-pzplot}, the blue diamond markers
indicate the sample zero locations and the magenta circle
markers indicate the UC MVDR zeros projected on unit
circle. \figurename{}~\ref{fig:smi-ucmvdr-bpplot} shows nulls and
lowered sidelobes in the UC MVDR beampattern (solid magenta) in contrast
to the shallow notches and higher sidelobes of the SMI MVDR
beampattern (dashed blue). 



\begin{algorithm}
  \caption{UC MVDR beamformer} \label{alg:ucmvdr}
  \begin{algorithmic}
    \Procedure{SMI MVDR}{$\datavec{}$}\Comment{Compute SMI MVDR weights}
     \State $\sampCov = \frac{1}{L}\sum\limits_{\ell=1}^L\datavec\datavec\herm$
     \State $\wsmi = {\sampCov\inv\replook}/{(\replook\herm\sampCov\inv\replook)}$
    \EndProcedure
    \Procedure{ProjectUnitCircle}{$\wsmi$}\Comment{Project zeros to unit circle}
    \State $\smipoly(z) = \ztrans (\wsmi\herm) = G\prod\limits_{n=1}^{N-1}(1 - \sampz_nz\inv)$ 
    \State  $\sampz_n = r_ne^{j\omega_n}$ \Comment{SMI MVDR polynomial zero}
     \If{$|\omega_n| > 2\pi/N$}
     \State $\ucz_n = e^{j\omega_n}$     
     \ElsIf{$|\omega_n| \leq 2\pi/N$}
     \State $\ucz_n = e^{\sign(\omega_n){j2\pi/N}}$
     \EndIf
     \EndProcedure
     \Procedure{UC MVDR}{$\ucz_n$}\Comment{Compute UC MVDR weights}
    \State $\ucpoly(z) = \prod\limits_{n=1}^{N-1}(1 - \ucz_n z\inv)/(1 - \ucz_n) = \sum\limits_{n=0}^{N-1} c_n^*z^{-n}$ 
    \State $\wuc = [c_1, c_2,\ldots,c_N]\trans$
    \EndProcedure
  \end{algorithmic}
\end{algorithm}

%% file: results.tex
\section{Simulation results}
\label{sec:ucbf-perf}
This section presents results of the simulation experiments evaluating
the performance of the UC MVDR beamformer compared to the SMI and the DL
MVDR beamfromers. All beamformers are implemented using both $N = 11$ and $N = 51$
element ULAs and steered to the broadside look
direction ($\ulook = 0$). The experiments assume a passive sonar
environment where the beamformers often operate with barely sufficient
snapshots $L = N + 1$ and two snapshots per sensor $L = 2N$ is
considered a snapshot rich scenario
\cite{cox2002adaptive,baggeroer1999passive}. The desired signal is not present in the snapshots used to estimate the beamformer weights. The assumption of desired signal-free snapshots is commonly used in beamformer design in practical sonar applications \cite{gershman2003robust}. In many passive sonar applications, the beamformers are continuously scanning the observation scene by steering the beampattern across the bearing range. When the beamformer beampattern main lobe is not steered towards any planewave source direction, it is appropriate to model the observed snapshots as signal free.

The simulated snapshots consist of a single loud interferer present at $\uinter = 3/N$ and a
unit power white background noise. For this measurement scenario, the
beamformer output power $\pout$ is the sum of the interferer contribution
$\pinter = \interfpow|\wt\herm\repint|^2$ and the noise contribution $\pnoise = \noisepow\norm{\wt}^{2}$. In the sequel, the two
components of the output power are referred to as the interferer
output power ($\pinter$) and white noise output power
($\pnoise$). \sect{}\ref{sec:ucmvdr-interf-outp-result} compares the
interferer output power of the beamformers and
\sect{}\ref{sec:ucmvdr-wng-result} compares the WNGs of the
beamformers. Since the white noise output power is determined by the WNG, it is sufficient to compare the beamformer's WNG. All the results are averaged from 3000 Monte Carlo trials.




To ensure a fair comparison between the UC MVDR beamformer and the DL MVDR
beamformer, the DL level ($\dl$) can be chosen to match either the average
WNGs or the average notch depths (ND) between the two beamformers. In most of
the experiments presented, the DL level is chosen to match the average
WNG between the UC MVDR and the DL MVDR beamformer. In order to determine the
DL level, the experiments first implement the UC MVDR for all trials
and compute the average WNG of the UC MVDR beamformer. The DL level is then estimated iteratively to match the average WNG between the UC MVDR beamformer and the DL MVDR beamformer. In an additional set of experiments, the UC MVDR and DL MVDR performances are compared using the diagonal loading that optimizes SINR per Mestre and Lagunas \cite{mestre2006finite}.

\subsection{Interferer output power}
\label{sec:ucmvdr-interf-outp-result}
\begin{figure}[tp]
  \centering
  \subfloat[$N = 11$, $L = 12$]{\includegraphics[width=3in]{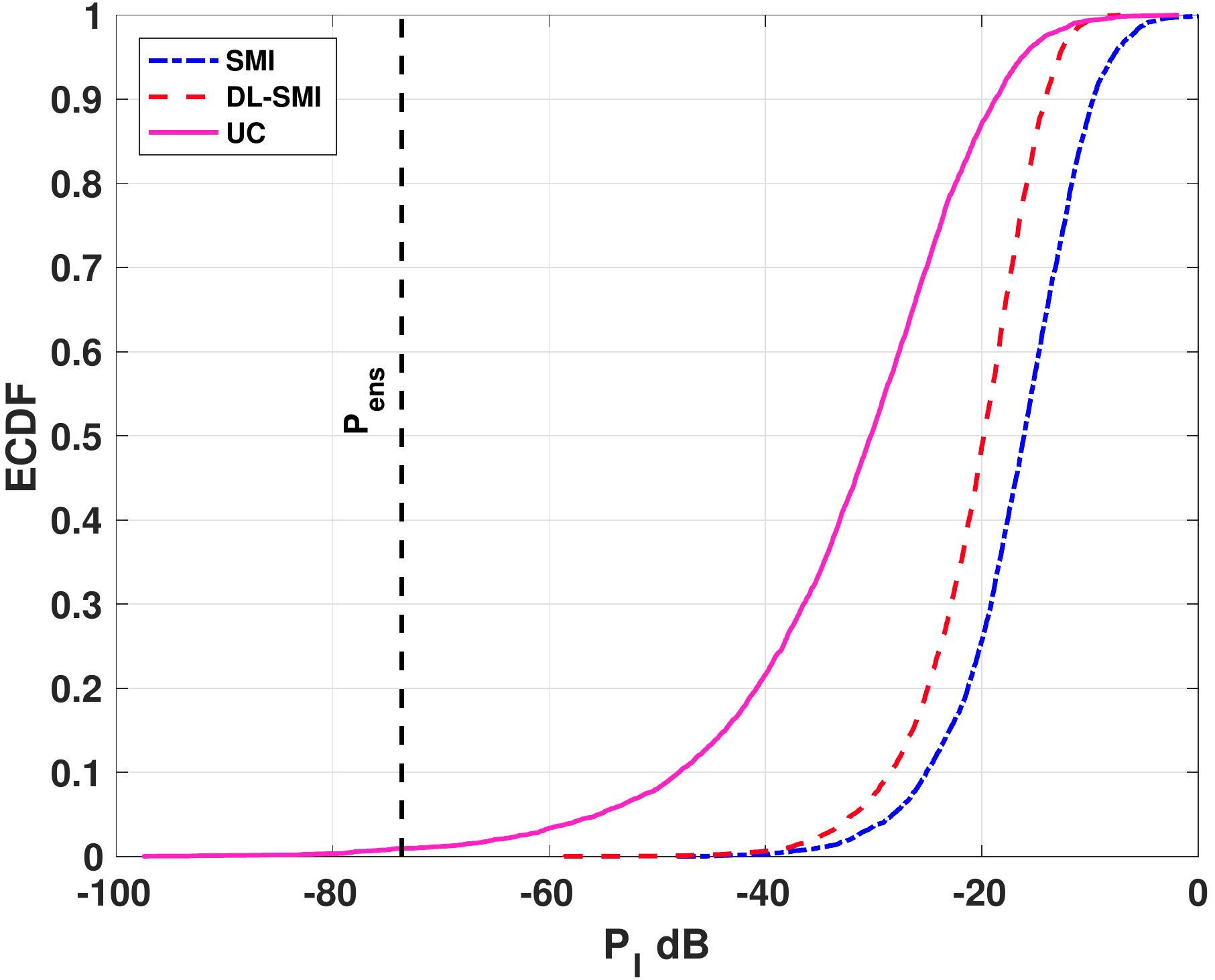}%
    \label{fig:ecdf_N11L12INR40}}
  \subfloat[$N = 11, L = 22$]{\includegraphics[width=3in]{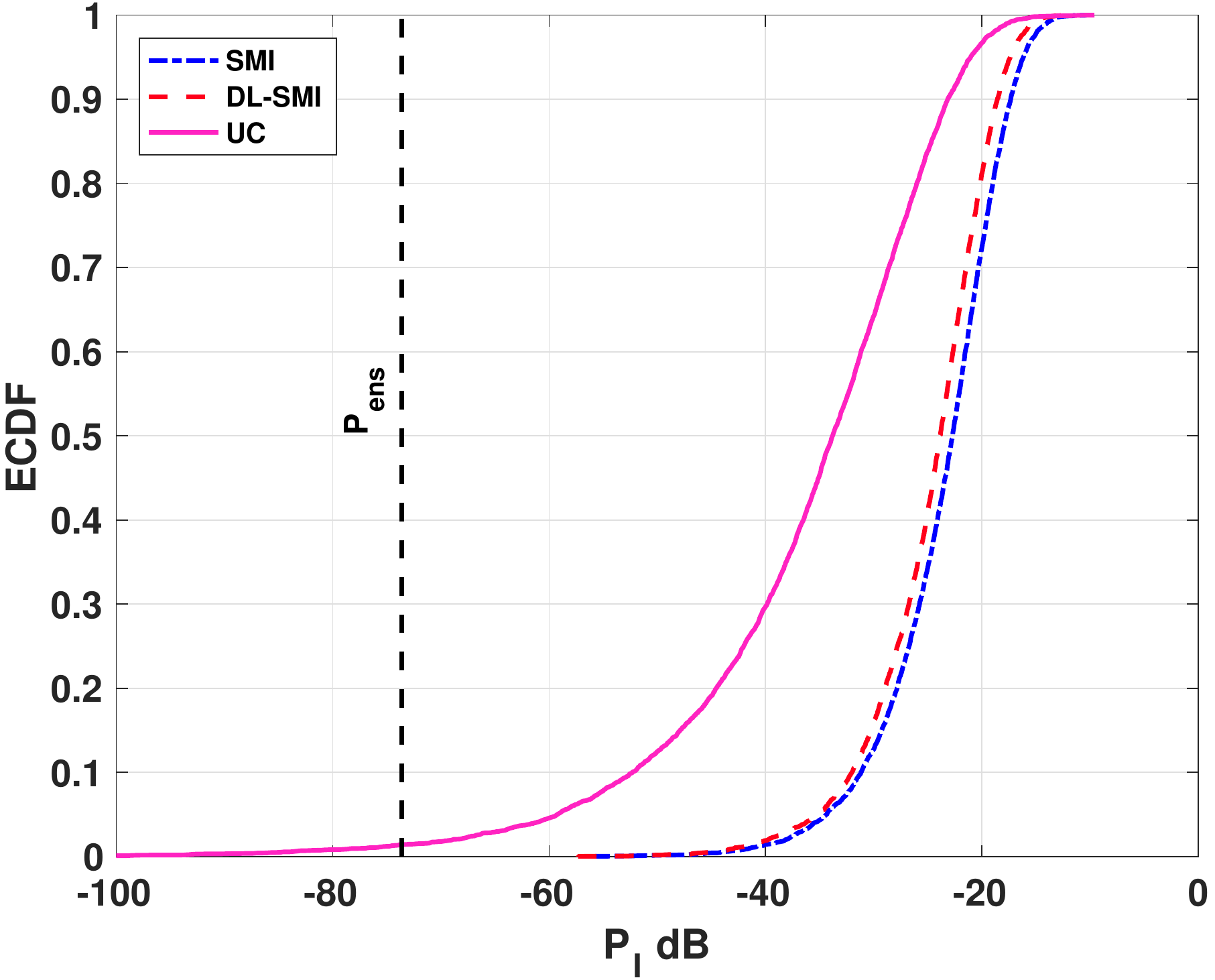}%
    \label{fig:ecdf_N11L22INR40}}\\
  \vfill
  \subfloat[$N = 51, L = 52$]{\includegraphics[width=3in]{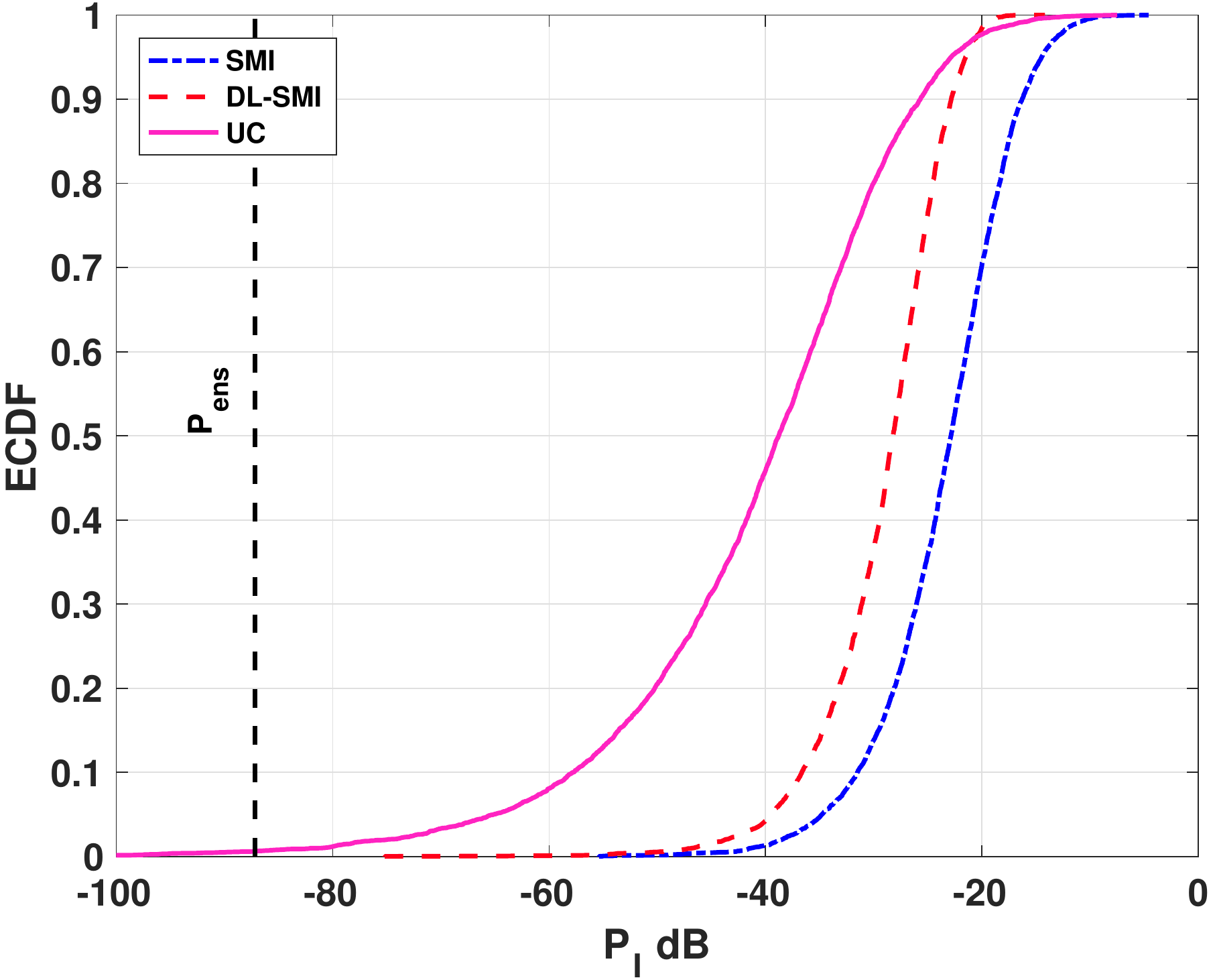}%
    \label{fig:ecdf_N51L52INR40}}
  \subfloat[$N = 51, L = 102$]{\includegraphics[width=3in]{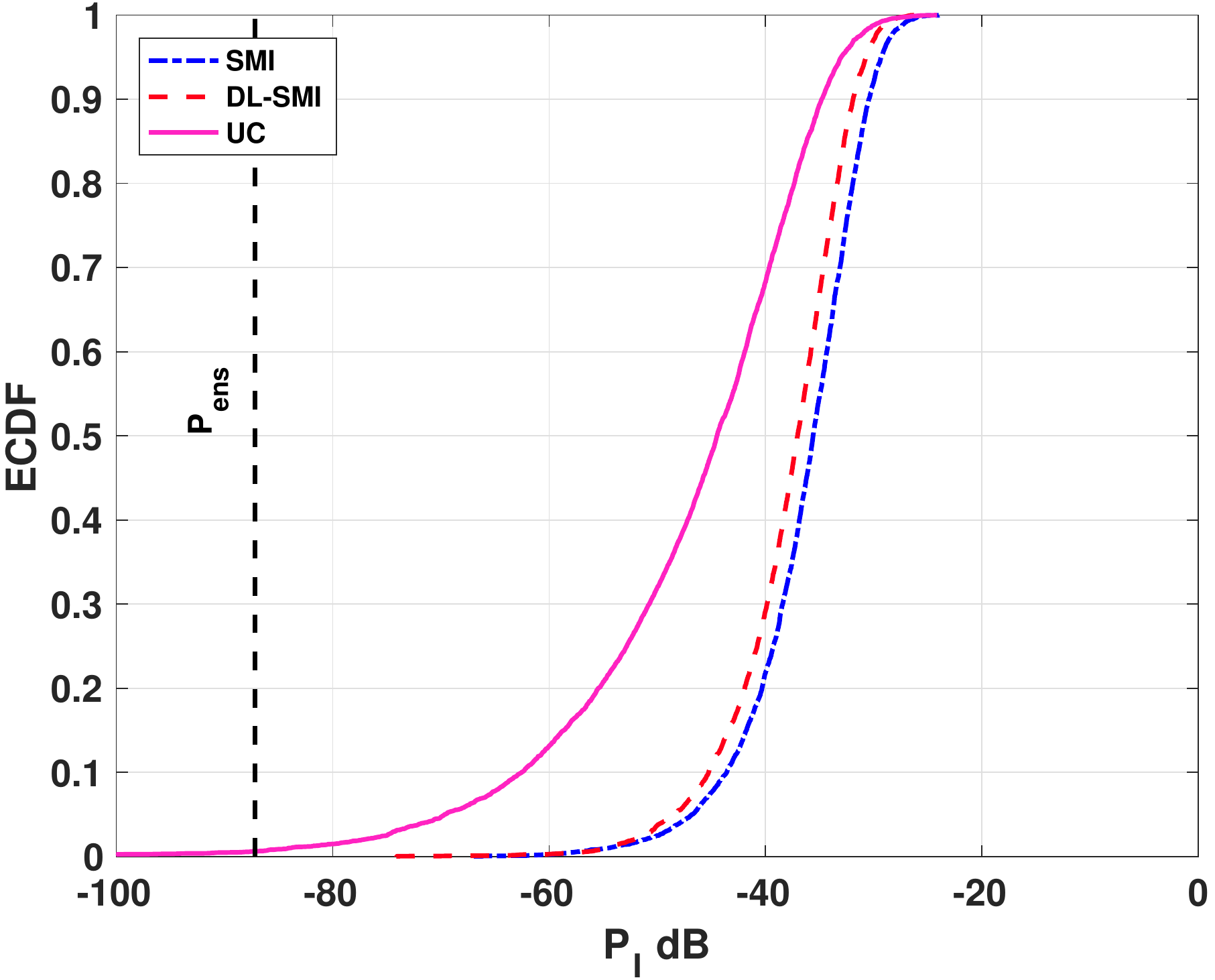}%
    \label{fig:ecdf_N51L102INR40}}
  \caption[ECDF of the interferer contributed output power $\pinter$
  for the SMI beamformer, the DL MVDR and the UC MVDR beamformer.]{ECDF of the interferer contributed output power $\pinter$ for the SMI beamformer, the DL MVDR and the UC MVDR beamformer. The beamformers use $N = 11$ element ULA in the top panel and $N = 51$ element ULAs in the bottom panel. The left panels shows the snapshot deficient case($L = N + 1$) and the right panels shows the snapshot rich case ($L = 2N$). In each panel the dashed vertical line denotes the interferer contributed output power for the ensemble MVDR beamformer. In all cases, the UC MVDR beamformer suppresses the interferer better compared to the SMI and DL MVDR beamformers.}
  \label{fig:ecdf-plots}
\end{figure}

\figurename{}~\ref{fig:ecdf-plots} shows the empirical cumulative
distribution function (ECDF) of the interferer output power
($\pinter$) for the UC MVDR beamformer compared to the SMI and DL MVDR
beamformers. The upper two panels show the ECDF graphs for the $N = 11$
sensor ULA and the lower two panels show the ECDF graphs for the
$N = 51$ sensor ULA. The left two panels show the limited snapshot
cases where $L = N + 1$ and the right two panels show the snapshot
rich case where $L = 2N$. For all ULA sizes and snapshot cases, the
sensor level INR was set at $40$ dB. The dashed vertical line
represents the ensemble interferer output power $P_{\rm ens}$ obtained
from the MVDR beamformer implemented using the ECM. The interferer
output power $\pinter$ corresponding to the ECDF equal to $0.5$
defines the median output power. The closer the median interferer
output power $\pinter$ of the beamformers is to the dashed vertical line, the
greater the probability of producing output comparable to the ensemble
case. For all four cases examined, the DL level for the DL MVDR beamformer is
chosen to match the average WNGs as described earlier.

Over the observed interferer output power range in
\figurename{}~\ref{fig:ecdf-plots}, the UC MVDR beamformer exhibits
higher probability of achieving lower interferer output power
$\pinter$ compared
against both SMI and DL MVDR beamformers. For instance in
\figurename{}~\ref{fig:ecdf_N11L12INR40} the median output power of UC
MVDR was approximately $14$~dB lower than SMI beamformer and $10$~dB lower
than DL MVDR beamformer. The DL MVDR beamformer has improved interferer
suppression over SMI beamformer as expected, but the UC MVDR beamformer has improved performance compared to both SMI and DL MVDR beamformers.

\begin{figure}[tp]
  \centering
  \subfloat[$L$ = 12 snapshots]
{\includegraphics[width=3.5in]{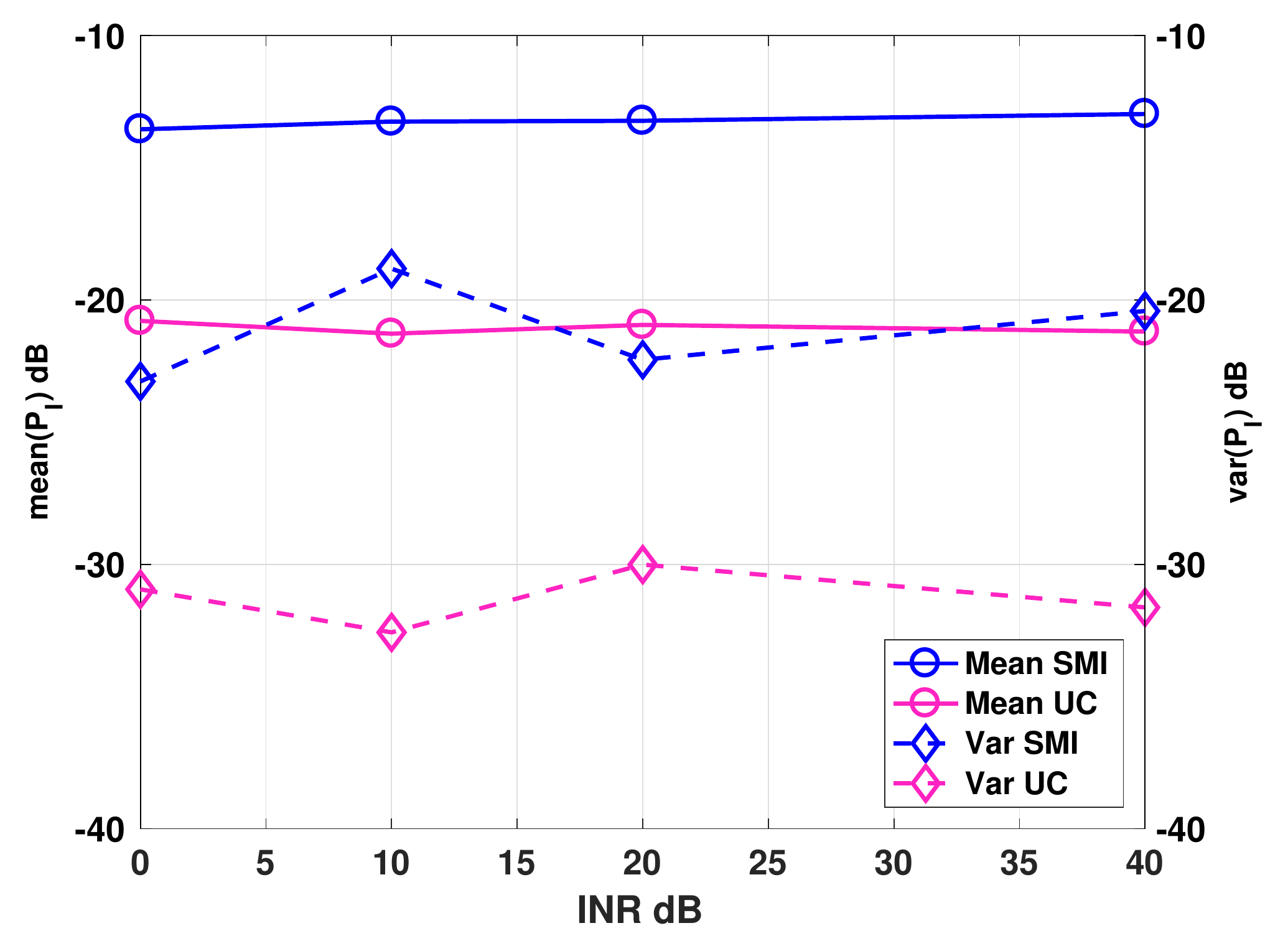}%
    \label{fig:mpout_N11L12}}  \hfill
  \subfloat[$L$ = 22 snapshots]
{\includegraphics[width=3.5in]{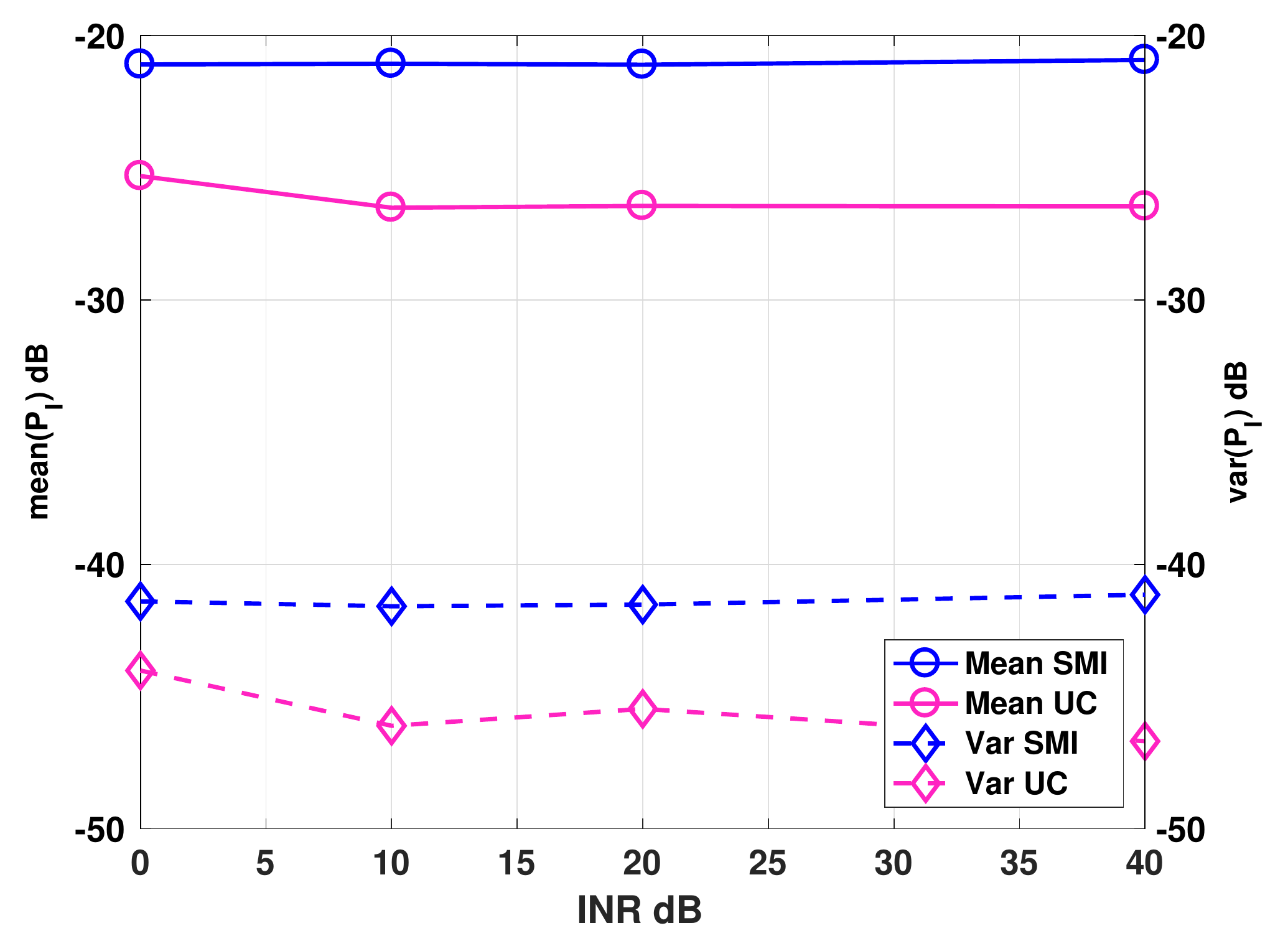}%
    \label{fig:mpout_N11L22}}
  \caption[Dual y-axis plot of mean and variance of interferer output power $\pinter$ for the SMI (blue) and the UC MVDR beamformer (magenta).]{Dual y-axis plot of mean and variance of interferer output power $\pinter$ for the SMI (blue) and the UC MVDR beamformer (magenta). Both beamformers are implement using $N = 11$ element ULA and SCM computed from $L = 12$ snapshots in the left panel and $L = 22$ snapshots in the right panel. The UC MVDR beamformer yields interferer contributed output power with lower mean and variance compared to the SMI beamformer.}
  \label{fig:pout-mean-var-plots}
\end{figure}

\figurename{}~\ref{fig:pout-mean-var-plots} compares the squared mean
(solid) and variance (dashed) of interferer output power $\pinter$ for
SMI and UC MVDR beamformers over a range of INR values from 0~dB to 40~dB. The interferer output power $\pinter$ of the UC MVDR beamformer has lower mean and variance compared to the interferer output power $\pinter$ of the SMI beamformer. The reduced mean interferer output power $\pinter$ when using the UC MVDR beamformer supports the earlier conclusion that the UC MVDR beamformer provides improved interferer suppression. 

\subsection{White noise gain}
\label{sec:ucmvdr-wng-result}
\begin{figure}[tp]
  \centering
  \subfloat[]{\includegraphics[width=3.3in]{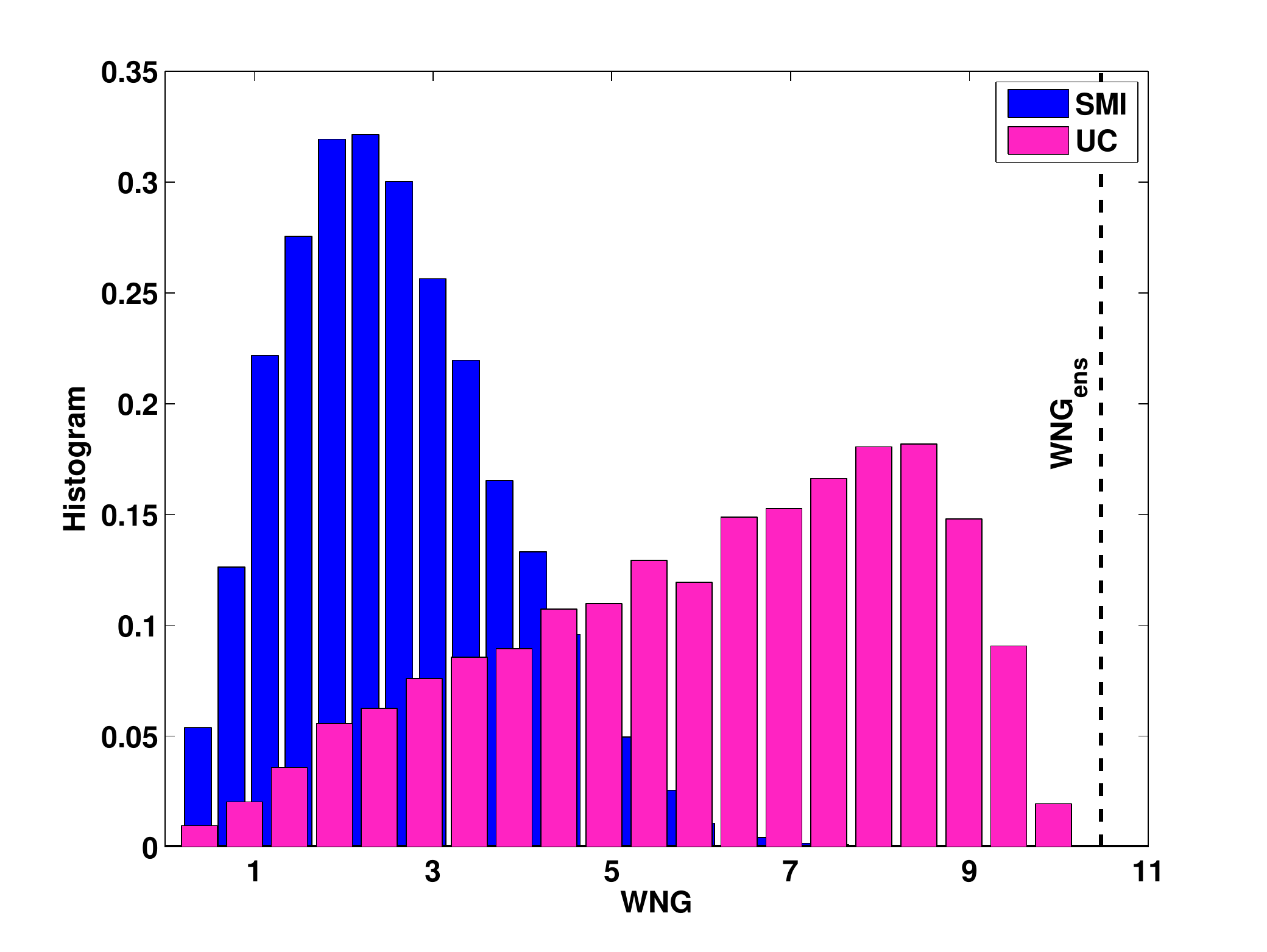}
    \label{fig:wng-hist-plot}}

\subfloat[]{\includegraphics[width=3in]{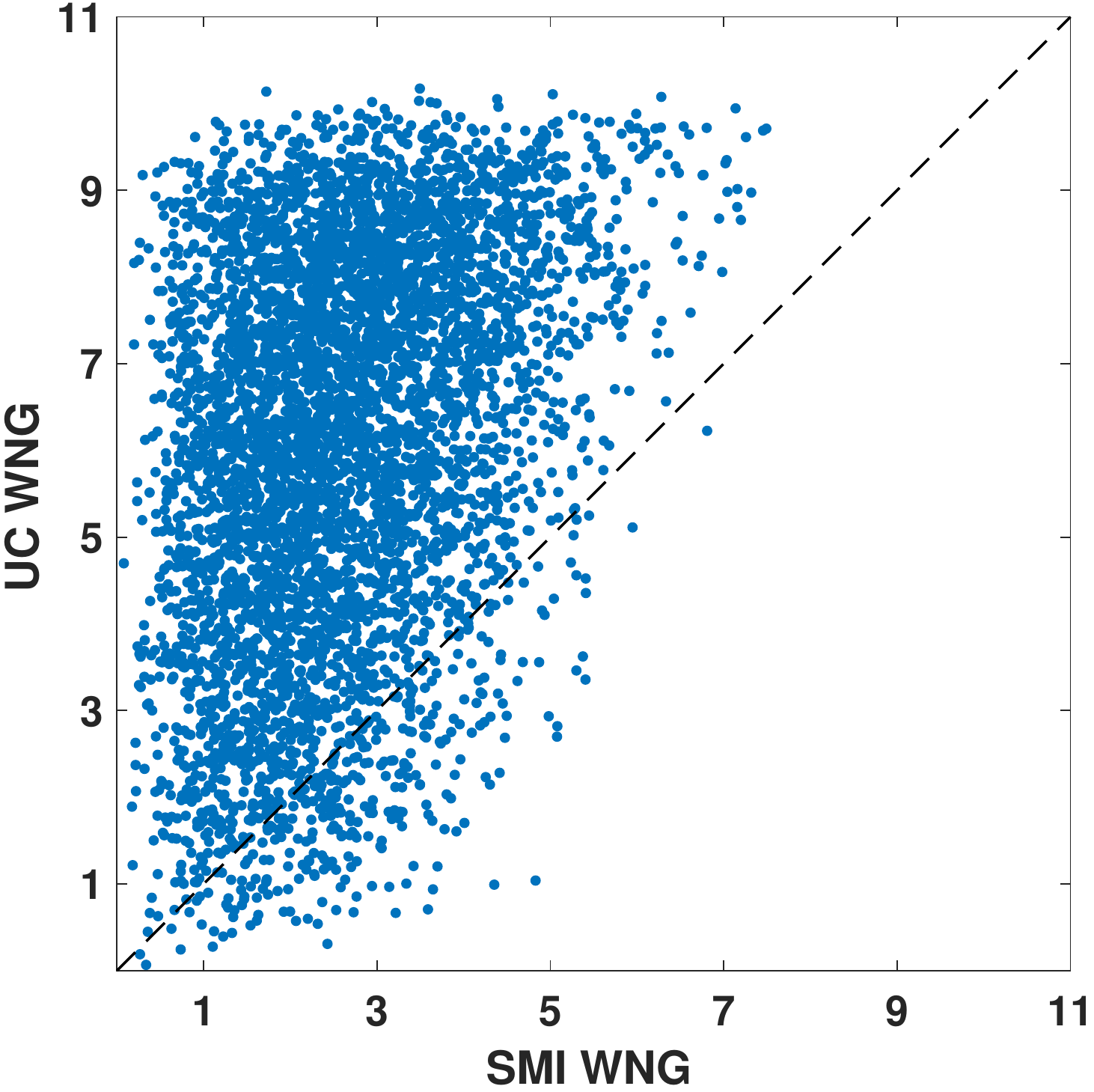}
    \label{fig:wng-scatter-plot}}
  \caption[Comparison of WNG for the UC MVDR and the SMI beamformer,
    both implemented using an $N = 11$ element ULA and $L = 12$ snapshots.]{Comparison of WNG for the UC MVDR and the SMI beamformer,
    both implemented using an $N = 11$ element ULA and $L = 12$
    snapshots. The top panel compares the histogram of WNGs and the lower
    panel is a scatter plot of the WNGs. On average, the UC MVDR beamformer
    has higher WNG compared to the SMI beamformer. }
  \label{fig:wng}
\end{figure}

\figurename{}~\ref{fig:wng} compares the WNG of the SMI and UC MVDR beamformers implemented using an $N = 11$ sensor ULA and $L = 12$
snapshots. The maximum possible WNG for this experiment is $WNG = 11$ and corresponds to the CBF \cite{vtree2002oap}. \figurename{}~\ref{fig:wng-hist-plot} compares the histograms of the WNG for the SMI and UC MVDR beamformers. The dashed vertical line denotes the ensemble MVDR WNG of $10.473$. The UC MVDR beamformer has a greater probability of achieving higher WNG with an average WNG of $5.672$ compared to an average WNG of $2.629$ using the SMI beamformer.

\figurename{}~\ref{fig:wng-scatter-plot} is a scatter plot comparing the WNGs
for the UC MVDR and the WNG for the SMI beamformers. Each point in the scatter plot denotes the WNG of the UC MVDR beamformer against the WNG of the SMI beamformer for a single realization of the beamformers in the Monte Carlo experiment. The scatter plot shows that the UC MVDR beamformer has a higher WNG than the SMI beamformer in most realizations except for small number of cases (bottom left corner in \figurename{}~\ref{fig:wng-scatter-plot}) where both beamformers have low WNG. Similar results were observed for the case of $N = 51$ sensor ULA. Thus projecting the sample zeros back to the unit circle frequently but not always improves the WNG from the SMI beamformer. As a result, the average WNG for the UC MVDR beamformer is better than that of the SMI beamformer for the cases examined here.



\subsection{UC MVDR and DL MVDR polynomial zeros}
\label{sec:ucmvdr-dlmvdr}
The UC MVDR and the DL MVDR beamformer are both derived by modifying the SMI beamformer. As described above, the UC MVDR beamformer moves the sample zeros radially back on to the unit circle. This section discusses how DL changes the DL MVDR zeros and compares this with the UC MVDR polynomial zeros.

\begin{figure}[tp]
  \centering
  \includegraphics[width=3.5in]{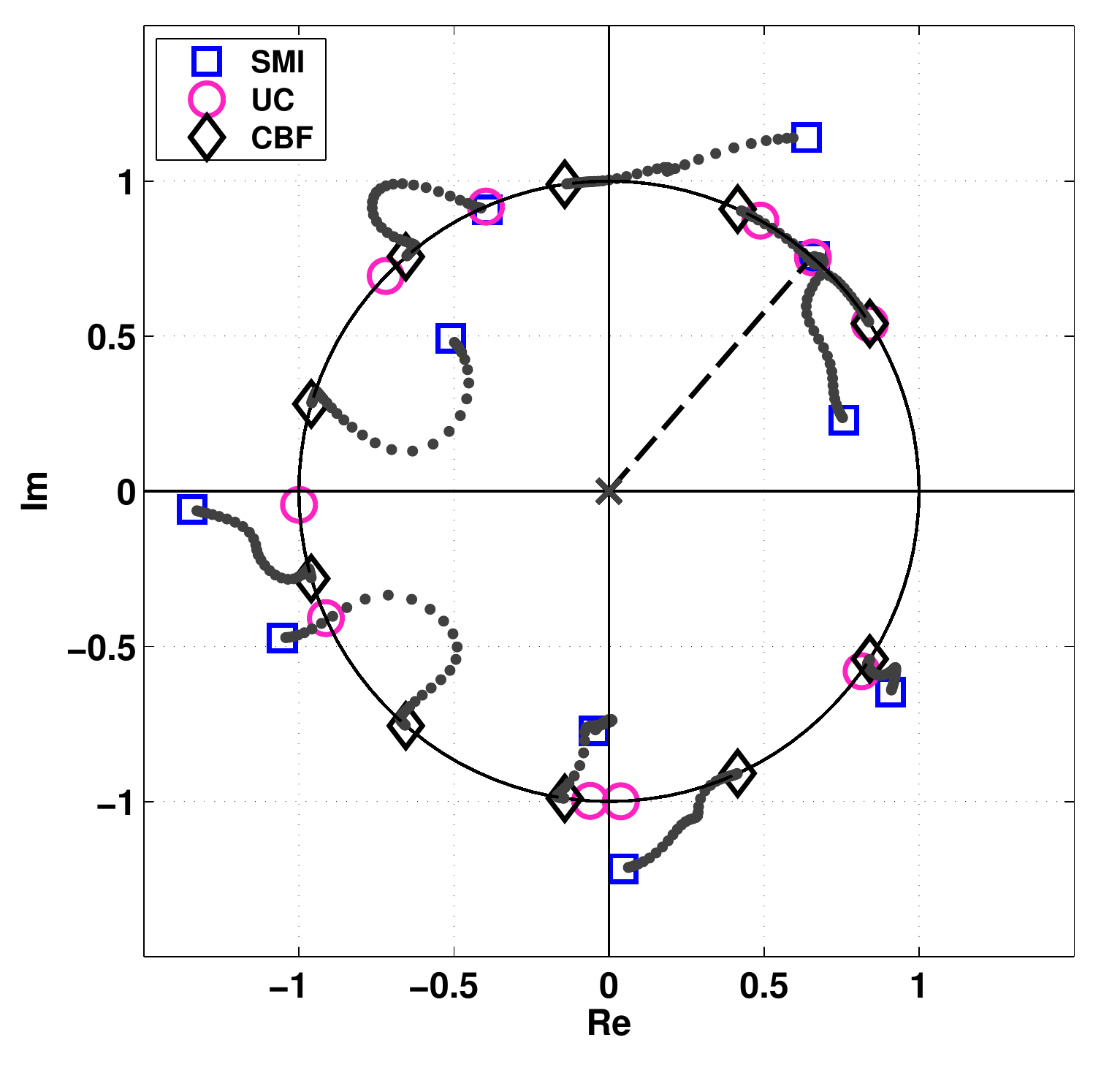}
  \caption[UC MVDR polynomial zeros compared against DL MVDR
    polynomial zeros as the DL level $\dl$ is increased from $-40$ dB to
    $40$ dB.]{UC MVDR polynomial zeros compared against DL MVDR
    polynomial zeros as the DL level $\dl$ is increased from $-40$ dB to
    $40$ dB. The DL MVDR polynomial zeros asymptotically converge to CBF polynomial zero locations as $\dl \rightarrow \infty$. }
  \label{fig:ucbf-dlsmi-pzplot}
\end{figure}

\figurename{}~\ref{fig:ucbf-dlsmi-pzplot} shows array polynomial zeros
for a representative example of beamformers implemented using an $N = 11$ element
ULA and steered to $\ulook = 0$. A single interferer is present at
$\uinter = 3/N$ denoted by the radial dashed line. The blue squares
denote the sample zeros, the magenta circles denote the UC MVDR zeros
and the black diamonds denote the CBF zeros. Each black dot denotes a
DL MVDR zero location as the DL level changes from $-40$ dB to $40$ dB
in $4$ dB steps. When the DL level $\dl = -40$~dB, the DL MVDR zeros
are essentially in the sample zero locations. As the DL level increases, the DL MVDR zeros converge towards the CBF zero locations as denoted
by the intermediate dot markers. The intermediate dot markers trace a
trajectory of DL MVDR zero locations starting from the sample zero
location to CBF zero location, as the DL level changes. A specific
trajectory is associated with each sample zero. Hence, changing the DL
level moves the DL MVDR zeros along specific trajectories. As
previously discussed in \sect{}\ref{sec:mvdr-beamformer}, Mestre and
Lagunas present an approach to compute the optimal DL level
\cite{mestre2006finite}. The DL MVDR zeros associated with the optimal
DL level are still constrained on the specific trajectories of each
sample zero, and may not be particularly close to the ensemble zero
locations. Comparatively, the UC MVDR beamformer uses a markedly different approach by moving the sample zeros radially back to the unit circle.

Moreover, one advantage of DL is that it improves the WNG of the beamformers
\cite{vtree2002oap}. As the DL MVDR zeros move closer to the CBF zero
locations, the WNG performance improves. However, moving zeros closer
to the CBF zero locations leads to loss of ND in the interferer
direction. Hence choosing the DL level involves a trade off between loss
of interferer suppression and improved WNG. By moving the zeros to the unit circle, the UC MVDR creates beampattern nulls, which improve the interferer suppression compared to DL MVDR, and as a by-product often improves WNG as well.
Further, the UC MVDR beamformer does not require choosing a tuning parameter like the DL
level.

\begin{figure}[t]
  \centering
  \includegraphics[width=3.5in]{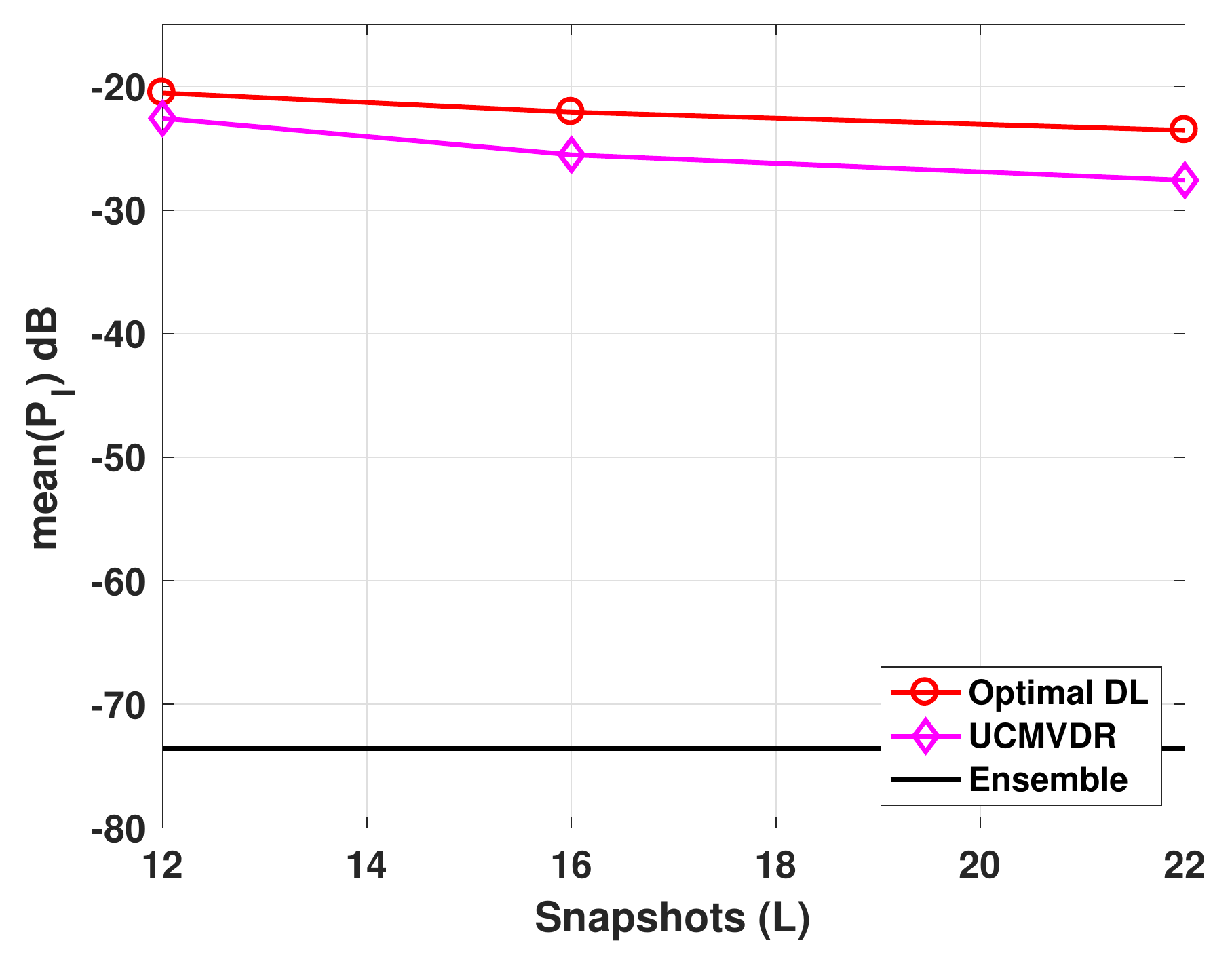}
  \caption{Interferer output power for the UC MVDR and the optimal DL MVDR beamformers for a range of snapshot values $12\leq L \leq 22$. The horizontal solid line denotes the interferer output power for the ensemble MVDR beamformer. The
beamformers are implemented using an $N = 11$ sensor ULA and steered towards
broadside look direction ($\ulook = 0$). A single interferer with INR
= $40$ dB is present at $\uinter = 3/N$. The UC MVDR beamformer suppresses the interferer better than the optimal DL MVDR beamformer whose DL level is computed using the knowledge of the ECM.}
  \label{fig:pout-uc-mestre}
\end{figure}

\figurename{}~\ref{fig:pout-uc-mestre} compares the mean interferer
contributed output power for the UC MVDR (magenta diamonds) and the
optimal DL MVDR (red circles) beamformers for a range of snapshot values
$12\leq L \leq 22$. The optimal DL level is computed using the approach
detailed in \cite{mestre2006finite}, assuming the knowledge of the ECM. The horizontal solid line denotes the interferer contributed output power
for the ensemble MVDR beamformer. The beamformers are implemented using an $N = 11$
sensor ULA and steered towards broadside look direction
($\ulook = 0$). A single interferer with INR = $40$~dB is present at
$\uinter = 3/N$. Comparing the output powers shows that the UC
MVDR beamformer suppresses the interferer better than optimal DL MVDR beamformer
over a range of snapshot values. As discussed earlier, applying DL
 moves the sample zeros away from the ensemble locations and
towards the CBF locations. As a result the optimal DL MVDR is improving
the WNG at the expense of interferer suppression. 

The improved interferer suppression of the UC MVDR beamformer in \fig\ref{fig:pout-uc-mestre} is impressive in light of the asymmetrical information between the beamformers in favor of the DL MVDR beamformer.  The DL MVDR beamformer has perfect knowledge of the ECM to compute the optimal DL, which is then used to process SCM data for the performance plotted in \fig\ref{fig:pout-uc-mestre}.  In contrast, the UC MVDR is operating only with SCM data. The UC MVDR beamformer rejects interferers better than the DL MVDR even when the latter beamformer has access to perfect information for optimizing the DL level.

%% file: discussion.tex
\section{Discussion}
\label{sec:discussion}
This paper seeks to identify the value of the UC rectification process, which is most clearly displayed in isolation separate from other rectification or regularization techniques. As such, the only main-lobe protection incorporated in the UC MVDR beamformer is the zero exclusion described in Sec.~\ref{sec:ucmvdr-algorithm}. In practice, the UC rectification would likely be combined with other covariance matrix and array weight conditioning techniques such as Toeplitz rectification \cite{cadzow1987,fuhrmann1991toeplitz,vallet2014}, diagonal loading \cite{vtree2002oap,ryan2017ucdlmvdr}, or limiting the white noise gain \cite{Cox1987robust}. Incorporating one or more of these techniques into our initial exploration of the algorithm might obscure the value of the UC rectification process. In isolating the UC rectification, we established the value of this technique in suppressing discrete interferers and often attenuating white noise as well. Future work can progress to more complicated issues such as the interactions of UC rectification with the covariance matrix conditioning techniques listed above. 

The proposed UC rectification algorithm excludes zeros within the null-to-null main-lobe width to prevent main-lobe distortion. This approach prevents the UC MVDR beamformer from suppressing discrete interferers inside the main-lobe. There are several promising candidates for protecting the main-lobe while suppressing main-lobe interferers using the UC MVDR beamformer. First, one could make the exclusion zone narrower around the look direction. Instead of excluding zeros within the null-to-null main-lobe width, one might only exclude them within the half power main-lobe width, sacrificing some WNG to suppress a loud interferer approaching the look direction. This approach is analogous to varying the inner product threshold in \cite{cox1997robust}. Another approach could introduce a constraint on the radius of the projected zeros within the main-lobe. The sample zeros within the main-lobe width would be radially projected towards the unit circle, but stop short of the unit circle based on a constraint on the WNG. This would create notches near the main-lobe interferers and produce smaller values of ND compared to the current formulation of the UC MVDR beamformer.

The UC MVDR beamformer uses radial projection as the simplest method for UC rectification comparable to Cadzow's the diagonal averaging approach as the simplest method to project the SCM to the nearest Toeplitz matrix \cite{cadzow1987}. Subsequent authors \cite{fuhrmann1991toeplitz,barton1997,vallet2014} developed more complicated maximum likelihood estimation algorithms for finding the most likely Toeplitz matrix, usually much more complicated than the diagonal averaging. We anticipate future refinements on the UC rectification exploiting progress establishing the distribution of the sample zeros as in \fig\ref{fig:smi-mvdr-pzplot} will likewise produce more sophisticated ML estimates of zeros projected onto the unit circle. However, the highly nonlinear and at times poorly conditioned nature of polynomial roots may make this more challenging than Toeplitz rectification \cite{cadzow1987,barton1997,vallet2014} or optimizing diagonal loading \cite{mestre2005diagonal, pajovic2014thesis}.

An important open challenge remains to find the probability distribution of the sample zeros perturbations away from the ensemble locations on the unit circle due to limited sample support and possibly also array element mismatch \cite{ryan2017thesis}. This is a challenging problem even for a second order polynomial, the distribution of the roots of quadratic as given by quadratic formula can be very complicated from the distributions of the polynomial coefficients. The complexity of this nonlinear mapping of probability distributions of the coefficients to probability distributions of the roots is beyond the scope of this paper, but when solved will resolve many of the open issues above. It would be a significant step to prove that the distribution of the perturbations of the zeros is circularly symmetric under some conditions.  The scatter plots of \fig\ref{fig:smi-mvdr-pzplot} tempt us to conjecture this must be the case for at least some problems, but we have not proven it at this point in time.  If the perturbation distribution is circularly symmetric, depending only on the radius of the perturbation and not the angle, the radial projection back to the unit circle may well prove to be a maximum likelihood estimate by minimizing the radius.


%

%% file: conclusion.tex
\section{Conclusion}
\label{sec:conclusion}
This paper proposed the UC MVDR beamformer which projects the SMI MVDR
array polynomial zeros radially to the unit circle to create perfect
notches in the beampattern. By moving the sample zeros to the unit
circle, the UC MVDR zeros satisfy the unit circle property on the
ensemble zeros. Numerical simulations verify that the UC MVDR beamformer often
simultaneously improves interferer suppression and WNG compared to the
SMI MVDR ABF. By moving the zeros onto the unit circle, the
UC MVDR is able to suppress interferers better than the optimal DL
MVDR whose DL factor is computed using perfect knowledge of the
ECM. Unlike DL MVDR, the UC MVDR does not require an \emph{a priori}
choice of parameter to achieve this improvement in performance.

%% file: appendix.tex
\section{Proof of unit circle property on ensemble MVDR array polynomial zeros}
\label{sec:apdx-mvdr-zeros}
The ensemble MVDR array polynomial zeros must be located on the unit circle
for planewave beamforming using ULA. Steinhardt and Guerci proved this
property of the MVDR ensemble zeros in \cite{steinhardt2004stap},
but the result does not appear to be widely known. The proof below
closely follows the Steinhardt and Guerci proof by contradiction.

The MVDR weight vector $\wmvdr$ solves the optimization problem in
\eqref{eq:mvdr-const-prob}. The quadratic objective function in
\eqref{eq:mvdr-const-prob} is a convex function in $\wt$ because the
ECM is a positive-definite matrix \cite{vtree2002oap}. This convexity implies a
unique solution exists for \eqref{eq:mvdr-const-prob}. The objective
function can be expressed in terms of the ensemble spatial power spectrum
$\powspec$ and MVDR beampattern

\begin{equation}
\label{eq:obj-func-spec}
f(\wt) = \half\int_{-1}^{1}\powspec |\beampatu|^2 \mathrm{d}u.
\end{equation}
Note that \eqref{eq:obj-func-spec} is a Parseval's relation equating the energy evaluated in the spatial frequency domain to the energy evaluated in the spatial
domain in \eqref{eq:mvdr-const-prob}.

Replace the beampattern with MVDR polynomial $\mvdrpoly{z = e^{j\pi u}}$ to get

\begin{equation}
\label{eq:obj-func-poly}
f(\wt) = \int_{-1}^{1}\powspec |\mvdrpoly{e^{j\pi u}}|^2 \mathrm{d}u.
\end{equation}
The zero locations correspond to the MVDR weights. Factor the polynomial for the solution of \eqref{eq:mvdr-const-prob} as

\begin{equation}
  \label{eq:mvdr-poly-zeros-format}
\mvdrpoly{e^{j\pi u}} = \prod\limits_{n=1}^{N-1}\frac{(1 - \ensz_i e^{-j\pi u})}{(1 - \ensz_i)}
\end{equation}
where $\ensz_i$s are the ensemble MVDR zeros and $\mvdrpoly{z} = 1$ for $z = 1$.
Using \eqref{eq:mvdr-poly-zeros-format} in \eqref{eq:obj-func-spec},
the objective function is

\begin{equation}
\label{eq:obj-func-zeros-form}
f(\wmvdr) = \half\int_{-1}^{1}\powspec \prod\limits_{n=1}^{N-1}\frac{(1 -
   \ensz_n e^{-j\pi u})}{(1 - \ensz_n)} \frac{(1 - \ensz_n^*e^{j\pi u})}{(1 - \ensz_n^*)}\mathrm{d}u
\end{equation}

Assume the zeros in \eqref{eq:mvdr-poly-zeros-format} are not on the
unit circle. Replacing one zero $\ensz_i$ by its conjugate-reciprocal
$1/\ensz_i^*$ in \eqref{eq:obj-func-zeros-form} leaves the objective
function unchanged. However, changing the zero alters the polynomial $\mvdrpoly{z}$ resulting in a different weight vector. This implies there exists a weight vector
different from the weight vector corresponding to
\eqref{eq:mvdr-poly-zeros-format} which still optimizes
\eqref{eq:mvdr-const-prob}. This contradicts with the uniqueness of
the solution to \eqref{eq:mvdr-const-prob}. The uniqueness holds only
when the zeros of $\mvdrpoly{z}$ are on the unit circle. Hence the MVDR
zeros must be on the unit circle.